\documentclass[a4paper,11pt]{article}
\pdfoutput=1 
\usepackage{jcappub}

\usepackage[T1]{fontenc}
\usepackage{xcolor}

\usepackage{comment}
\newcommand{\be}{\nopagebreak[3]\begin{equation}}
\newcommand{\ee}{\end{equation}}

\title{\boldmath Constraints on unimodular diffusion models with latest observables}

\author[a,1]{Chavarr\'ia Jes\'us,\note{Corresponding author.}}
\author[b,c]{Fromenteau Sebastien,}
\author[a]{Sudarsky Daniel,}
\author[d]{Vargas-Magaña Mariana}

\affiliation[a]{Instituto de Ciencias Nucleares, Universidad Nacional Aut\'{o}noma de M\'{e}xico, 04510, Cd. Mx., M\'{e}xico.}

\affiliation[b]{Instituto de Ciencias F\'{i}sicas, Universidad Nacional Aut\'{o}noma de M\'{e}xico, 62210, Cuernavaca, Morelos.}

\affiliation[c]{Instituto de Astronom\'{i}a, Universidad Nacional Aut\'{o}noma de M\'{e}xico, Apdo. 70–264, Cd. Mx., M\'{e}xico.}

\affiliation[d]{Instituto de F\'{i}sica, Universidad Nacional Aut\'{o}noma de M\'{e}xico, Apdo. Postal 20-364, 01000, Cd. Mx., M\'{e}xico.}

\emailAdd{jesus.soto@correo.nucleares.unam.mx}
\emailAdd{sfroment@icf.unam.mx}
\emailAdd{sudarsky@nucleares.unam.mx}
\emailAdd{mmaganav@fisica.unam.mx}

\abstract{Cosmological models incorporating a time-dependent equation of state have recently been explored \cite{DESI:2025fii}, showing a preference for a dynamical dark energy component. In this work, we investigate a scenario in which an effective, time-dependent cosmological constant arises as an emergent manifestation of a violation of energy–momentum conservation. 
In \cite{Landau:2022mhm}, such a violation of energy conservation was studied as a diffusion mechanism affecting matter (dark and baryonic), leading to an effective dark energy component within the framework of unimodular gravity. Here, we present an updated analysis using the more recent Type Ia supernova data set from the Dark Energy Survey (DESY5) and the baryon acoustic oscillation (BAO) measurements from the Dark Energy Spectroscopic Instrument (DESI) Data Release 2 (DR2), along with the CMB temperature, polarization, and lensing data from Planck 2018. 
Our results identify a transition phase that occurs at intermediate times, with slight evidence in favor of the model relative to the $\Lambda$CDM according to the $\mathrm{\Delta DIC}$ criterion. Interestingly, a non-decisive preference for an evolution corresponding to either a time-decreasing or time-increasing effective cosmological constant is found. However, slightly higher values of $H_0$ favor a time-increasing effective cosmological constant. Although the $H_0$ tension is not significantly alleviated, these results suggest that a more refined modeling of the physics of the diffusion mechanism may offer a viable route toward addressing the current discrepancy in the Hubble expansion rate, while also providing a natural framework for incorporating a dynamical dark energy and addressing the problem of vacuum energy contribution.}

\begin{document}
\maketitle
\flushbottom

\section{Introduction}
Cosmology has made significant progress with the release of data from surveys such as DESI DR2 and DES Y5. We now find ourselves in the era of high-precision cosmology.  The associated cosmological data have enabled stringent tests of the $\mathrm{\Lambda CDM}$ model and provided an opportunity to search for new physics. Although  $\mathrm{\Lambda CDM}$ is currently the most widely accepted cosmological model, it is not without issues. To be consistent with observations, it incorporates a dark energy component. This may initially seem unproblematic, since the standard Einstein field equations admit the addition of a cosmological constant, which can, in principle, be interpreted either as part of the spacetime geometry or as a form of matter–energy content. Conceptually, one might want to determine whether this term should be placed on the geometric or matter side of the equations.  When treated as a matter component, one finds that $\sim 70\%$ of the universe's energy budget is attributed to this mysterious, gravitationally repulsive substance. In this direction, various theoretical alternatives have been proposed to account for this component, including contributions from ultra-light axions, black holes, or a large deviation from homogeneity that could give rise to an effective back-reaction effect \cite{Ishibashi:2005sj, Christiansen:2024kxn}. Despite these efforts, the physical nature of dark energy remains elusive. On the other hand, and as is well known, 
recent results have highlighted and exacerbated existing problems, suggesting potential inconsistencies in the standard cosmological model \cite{DES:2025bxy, DESI:2025fii}, pointing to a time-dependent dark energy that better fits the observations. It is clear that, if one trusts this result, the cosmological constant can no longer be interpreted as a purely geometric term within standard general relativity.

Cosmological models that incorporate a time-dependent dark energy have been highly proposed \cite{Nagpal:2025omq}, two-parameter parametrization models like Chevallier-Polarski-Linder (CPL) \cite{Chevallier:2000qy}, Barboza-Alcaniz (BA) \cite{Barboza:2008rh}, JBP \cite{Jassal:2004ej} and logarithmic parametrization \cite{Wetterich:2004pv, Ma:2011nc, Efstathiou:1999tm} have been explored, showing a preference by more than $3\sigma$ over $\rm\Lambda CDM$ model. The problem with these parametric models is that they sometimes lack a phenomenological origin. They directly attribute the discrepancy with observations to an energy component through a parameterized function $\omega(a)$. These parametrization schemes also present the additional problem of crossing the phantom divide ($\omega < -1$), where the dominant energy is violated, which implies the existence of a tachyonic particle. Two possible approaches can be taken. One is to treat these parametrized models as an effective description of a dark energy component, without concern for the underlying physical nature of the proposal. Alternatively, one may motivate a model in which the dynamical behavior of dark energy arises from fundamental principles not associated with new cosmological energy components, in contrast to scalar-field models commonly considered in the literature \cite{1988ApJ...325L..17P, Shajib:2025tpd}.

At this point, we should point to the proposal made in \cite{Josset:2016vrq,PhysRevLett.118.021102, 2017arXiv171105183P}, where something that acts as an effective cosmological constant can emerge 'naturally' by considering the violation of the energy-momentum tensor in the context of unimodular gravity. Posterior refinements of these ideas involve considering physical mechanisms responsible for the violation of energy–momentum conservation, where it is taken as tied to a fundamental granularity of space-time, generating a diffusion-like effect in the evolution of ordinary matter in the very early post-inflationary cosmological epochs \cite{Perez:2019gyd}. Such consideration yields an estimate of the cosmological constant that is compatible with observations, in contrast to traditional estimates based on quantum-vacuum contributions, which differ by about 120 orders of magnitude.  Following the proposal in \cite{Perez:2020cwa}, we explore whether a process such as that discussed in \cite{Perez:2019gyd}, involving a diffusion-like effect associated with black holes, could generate an effective contribution in the post-decoupling epoch. Through the complete modeling of these ideas would involve not just the detailed form of the interaction of the spacetime granularity with individual black holes (something that has been schematically done in \cite{Perez:2020cwa}) but also the full cosmological evolution of the black hole distribution as a function of their mass and angular momentum, a task that we hope to address in future works. In the present manuscript we will instead follow the approach used in \cite{Perez:2020cwa} and \cite{Landau:2022mhm},  and instead of modeling all aspects of the idea in some detail,  we limit ourselves to a simple characterization of the behavior of the ordinary matter and dark energy components as a function of the cosmological time\footnote{As the reader can note, we use the terms effective cosmological constant and dark energy interchangeably. }.

Unimodular gravity (UG) is a theory explored by Einstein soon after he proposed the general theory of relativity in 1915 \cite{Einstein:1919gv}. The conceptual importance of the theory lies in that it offers a possible solution to the problem of the contribution of vacuum energy density from quantum matter fields in the semiclassical scenario, where quantum fields couple to the Einstein equations through their expectation values \cite{Ellis:2010uc}. This is the 120 order of magnitude difference between the predicted and observed values \cite{Riess:2019cxk}. By providing a natural framework in which vacuum energy does not gravitate, the theory offers a potential resolution of this problem.

In recent works, UG has also been proposed as a natural framework to account for the observed value of the cosmological constant and its apparent dynamical nature, as suggested by recent observations. In this framework, the cosmological constant is not linked to the energy content but instead emerges as an integration constant when the additional condition of energy-momentum conservation ($\nabla \cdot T = 0$) is imposed. Nevertheless, this condition does not necessarily need to be adopted. Unimodular gravity admits a limited kind of violations of the energy-momentum tensor (see discussion in \cite{Josset:2016vrq}). General relativity requires the conservation of the energy-momentum tensor, a result that emerges from the diffeomorphism invariance of the theory and the fact that fields satisfy classical equations. In unimodular gravity, this condition is relaxed as is appropriate for a situation in which the invariance is reduced to the volume-preserving diffeomorphisms\footnote{Considering a general infinitesimal diffeomorphism generated by a vector field $\xi$ acting on the matter action, the requirement that all fields be dynamical implies $\nabla \cdot T = 0$. Restricting to volume-preserving diffeomorphisms, i.e., those satisfying $\nabla \cdot \xi = 0$, instead leads to the constraint $d\mathbf{J} = 0$.}. Hence, UG reproduces the equations of general relativity with a cosmological constant in the case of a conserved energy–momentum tensor, while allowing for an effective, time-dependent cosmological term when $\nabla \cdot T \neq 0$.

The failure of a conserved energy-momentum tensor is quantified in terms of the current $J_a= 8\pi G \, \nabla^{b} T_{ab}$. In the cosmological context, $J=\dot{\Lambda}(t) \, dt$ due to spatial homogeneity and isotropy, $\Lambda(t)$ emerges as the effective cosmological constant which is generated by the cumulative effect of energy-momentum violation. So that  
\be
 \Lambda(x)=  \Lambda_0+\int_l J  
\label{eq. Effective_cosmological_constant}
\ee
is the effective cosmological constant. 

The field equations in unimodular gravity can be written in the form

\be
    R_{ab}-\dfrac{1}{2}g_{ab}R+\underbrace{\left(\Lambda_0+\int_l J\right)}_{\Lambda(x)}g_{ab} = 8\pi GT_{ab},
\label{Field Equations}
\ee
where $\Lambda_0$ is a constant,  representing the universe's initial conditions, $J_a$ is the current of energy-momentum violation, and $l$ is any arbitrary path connecting the "beginning of the universe" with the hypersurface of the corresponding fixed value of cosmological time. These equations are also known as the trace-free Einstein equations, obtaining by substituting $R_{ab}\rightarrow R_{ab}-\frac{1}{4}Rg_{ab}$ and $T_{ab}\rightarrow T_{ab}-\frac{1}{4}Tg_{ab}$ into the standard Einstein equations. An important consequence is that the theory naturally evades the contribution of vacuum energy to the gravitational dynamics, since matter contributes only through the trace-free part of $T_{ab}$ and any term proportional to the metric tensor vanishes identically.

Considering a homogeneous and isotropic universe described by the Friedmann-Lemaître-Robertson–Walker (FLRW) metric, the Friedmann and continuity equations in unimodular gravity take the form
\be
    H^2=\dfrac{8\pi G}{3} \sum_i \rho_i+\dfrac{\Lambda(t)}{3}, \quad \quad \dot{\rho}_m+3H \rho_m=- \dfrac{\dot{\Lambda}(t)}{8\pi G},
    \label{eq. Friedmann and cont}
\ee
where $\rho_m\equiv \rho_b+\rho_{c}$ is the energy density of baryonic matter and dark matter.

Concerning the physical mechanisms behind the violation of energy conservation that contribute to $J$, several natural possibilities can be identified. First comes from considering effects related to considering the quantum nature of matter in the context of objective collapse theories, for instance, within the continuous spontaneous localization (CSL) model, which leads to $J = -\xi_{\mathrm{CSL}} \rho^b dt$, as shown in \cite{Josset:2016vrq}. In fact, as argued in \cite{Maudlin:2019bje}, similar expectations apply more broadly to all reasonable paths for dealing with the measurement problem in quantum theory. 
Another context, explored by one of the collaborators \cite{Perez:2020cwa}, arises in connection with a possible manifestation of spacetime granularity through its influence on the matter propagating through it. This effect is  expected to become relevant in regions of high scalar curvature $\mathbf{R}$, and has been conjectured to be describable in terms of  modifications to the particle geodesic equation

\be
    u^{\mu} \nabla_\mu u^v = \alpha \dfrac{M}{m^2_p}\mathrm{sign} (s \cdot \xi) \mathbf{R} s^v,
    \label{Geodesiceq}
\ee
where $\alpha$ is a dimensionless coupling constant, $u$ is the four velocity of the particle, $s$ is the spin of the particle, $M$ the particle mass, $m_p$ the Planck mass, and $\xi$ a preferred time-like vector field characterizing the mean state of motion (in the cosmological case, the four velocity of the commoving observers). This equation is meant to be applied to fundamental particles and was the basis for the estimation of the value of the cosmological constant in \cite {2017arXiv171105183P}. On the other hand, when considering black holes, which, as discussed in \cite{2017arXiv171105183P}, in a suitable sense can be regarded as very similar to particles\footnote{In the sense that they are described by a few quantities: the charges at infinity (mass, angular momentum, and electric charge).}. It is possible to postulate an expression of the form \eqref{Geodesiceq} to describe black hole interaction with granularity, substituting  $\alpha \rightarrow \bar{\alpha}_{\mathrm{bh}}$ representing a new coupling constant and $\mathbf{R} \rightarrow \widetilde{\mathbf{R}}$ denoting an averaged notion of spacetime curvature associated with translational friction and, consequently, with matter that generates the mean curvature of spacetime, without including the proper curvature associate with the black-hole (see ref. \cite{Perez:2019gyd}). In this case, one must consider an equation for the black hole spin simultaneously, keeping in mind that $s \cdot u = 0$, and hence $\nabla (s \cdot u) = 0$, which leads to an expression for $u^\mu \nabla_\mu s^\nu$. Incorporating a spin diffusion term, one is led to postulate: 

\begin{align}
        u^{\mu} \nabla_\mu s^v = \bar{\alpha}_{\mathrm{bh}} \dfrac{M}{m^2_p}\mathrm{sign} (s \cdot \xi) \widetilde{\mathbf{R}}(s\cdot s) u^v- \bar{\beta}_{\mathrm{bh}} \dfrac{M}{m^2_p} \widetilde{\mathbf{R}}_{\mathrm{BH}} s^v,
    \label{Spin}
\end{align}
with $\widetilde{\mathbf{R}}_{\mathrm{BH}}$ the average local curvature around the black hole. Thus, as a consequence of Planck-scale granularity, such modifications could affect black hole dynamics, potentially allowing for changes in black hole spin. For a Kerr black hole, with mass $M$, there is about  $30\%$ energy density that can dissipate without violating the second law. Hence, following the previous discussion, a possible diffusion mechanism from the matter sector (e.g., involving black holes) to dark energy, mediated by a back-reaction associated either with the quantum nature of matter or with a black hole–granularity interaction, can be justified,  and treated within the UG framework. Meanwhile, the diffusion-like effect generated by fundamental particles, for suitable values of $\alpha$, has been shown to produce a cosmological constant at early times consistent with the observed one. However, no subsequent dynamical evolution is expected from particle translational energy diffusion (see ref. \cite{Perez:2019gyd}).  In this scenario, black holes provide a mechanism for generating a possible late-time cosmological constant.

Under this framework, we explore the nature of the diffusion process in an extended version of the previous analysis \cite{Landau:2022mhm}. In this context, a cosmological constant can be generated at late times through mechanisms such as continuous spontaneous localization, friction-like effects on particle propagation, or black hole spin energy diffusion. Modeling such effects as a simple diffusion process involving matter and dark energy, as a direct consequence of the violation of energy conservation, we consider a simple function describing dark energy generation, and investigate dark energy generation during the post-recombination epoch. In the present work,  we constrain the discrete model of \cite{Landau:2022mhm} and a continuous version of the previous model, using the most recent BAO and type Ia supernova data, combined with CMB from Planck. We also consider an extended version of the continuous model that allows for the possibility of a universe with non-vanishing spatial curvature. 

The paper is organized as follows. In Section~\ref{sec:Methodology and dataset}, we describe the methodology followed in this work.  For completeness, we describe both models and their respective parameterizations. We review the discrete model from previous analyses, presenting the equations for $\rho_b$ and $\rho_c$, and derive the corresponding continuous version by considering a flat Universe, incorporating a diffusion term of the form $\rho_\Lambda \sim \arctan(a)$. In Section~\ref{sec:Results}, we present the results of the parameter inference for both cases, as well as for two extended versions that include spatial curvature and the imposition of a prior on one of the unimodular parameters. In section \ref{sec:Appendix A}, the consistency of individual observables is tested by quantifying the tension in $\Omega_m$ and $H_0$. Section~\ref{sec: Appendix B} presents a comparison of our model with the standard parametrization schemes in the literature. Finally, Sections~\ref{sec:Discussion} and~\ref{sec:Conclusions} contain the discussion and conclusions, respectively.

\section{ Methodology}
\label{sec:Methodology and dataset}
\begin{figure}
\centering

\includegraphics[width=1\textwidth]{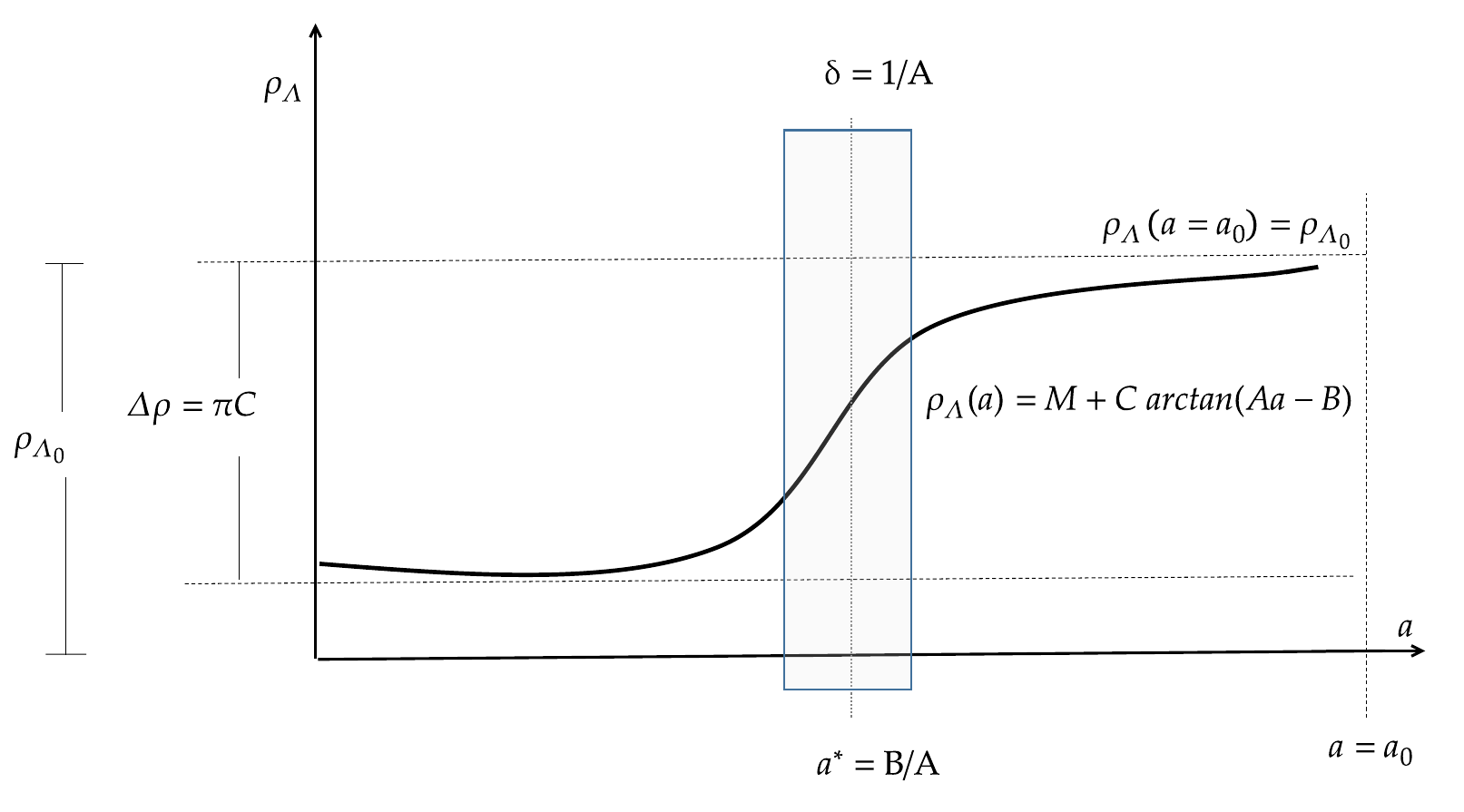}
\caption{Graphical representation of the continuous model, where the parameters A, B, M, and C can be expressed as functions of $\Delta \rho$, $a^*$, $\delta$, and $\rho_{\Lambda_0}$}
\label{Continous_model}
\end{figure}

To explore the possibility of time-dependent dark energy arising from a violation of energy–momentum conservation, we implemented two models in the Cosmic Linear Anisotropy Solving System (CLASS) \cite{2011arXiv1104.2932L, 2011JCAP...07..034B}. The first is a discrete model, based on step-like transitions and previously used in \cite{Landau:2022mhm}, illustrated in figure~\ref{Discrete_model}. The second is a smooth model, involving a continuous energy flow between matter and dark energy, also shown in figure~\ref{Continous_model}. In this section, we present the equations corresponding to each model. Afterwards, we show the resulting CMB Temperature and Polarization Power Spectrum. Finally, we present the data sets used in this analysis and other analysis choices.

\subsection{Discrete model}
\label{sec:Discrete model}
\label{theory1}
In \cite{Landau:2022mhm}, for homogeneous and isotropic Friedmann-Lema\^{\i}tre-Robertson-Walker (FLRW) universe, the non-energy conservation effect was modeled as diffusion process of matter (baryonic and dark matter) into an effective dark energy term given by a three-parameter expression for $\rho_\Lambda $, where only the matter degrees of freedom participate in the diffusion process\footnote{In a more general analysis, a diffusion process could also involve, for example, relativistic degrees of freedom in the context of the causal set approach to quantum gravity \cite{Philpott:2008vd}.}. This is motivated by the black hole diffusion mechanism mentioned previously and by the effects arising from considering the quantum nature of matter. 

\begin{figure} 
\centering

\includegraphics[width=1\textwidth]{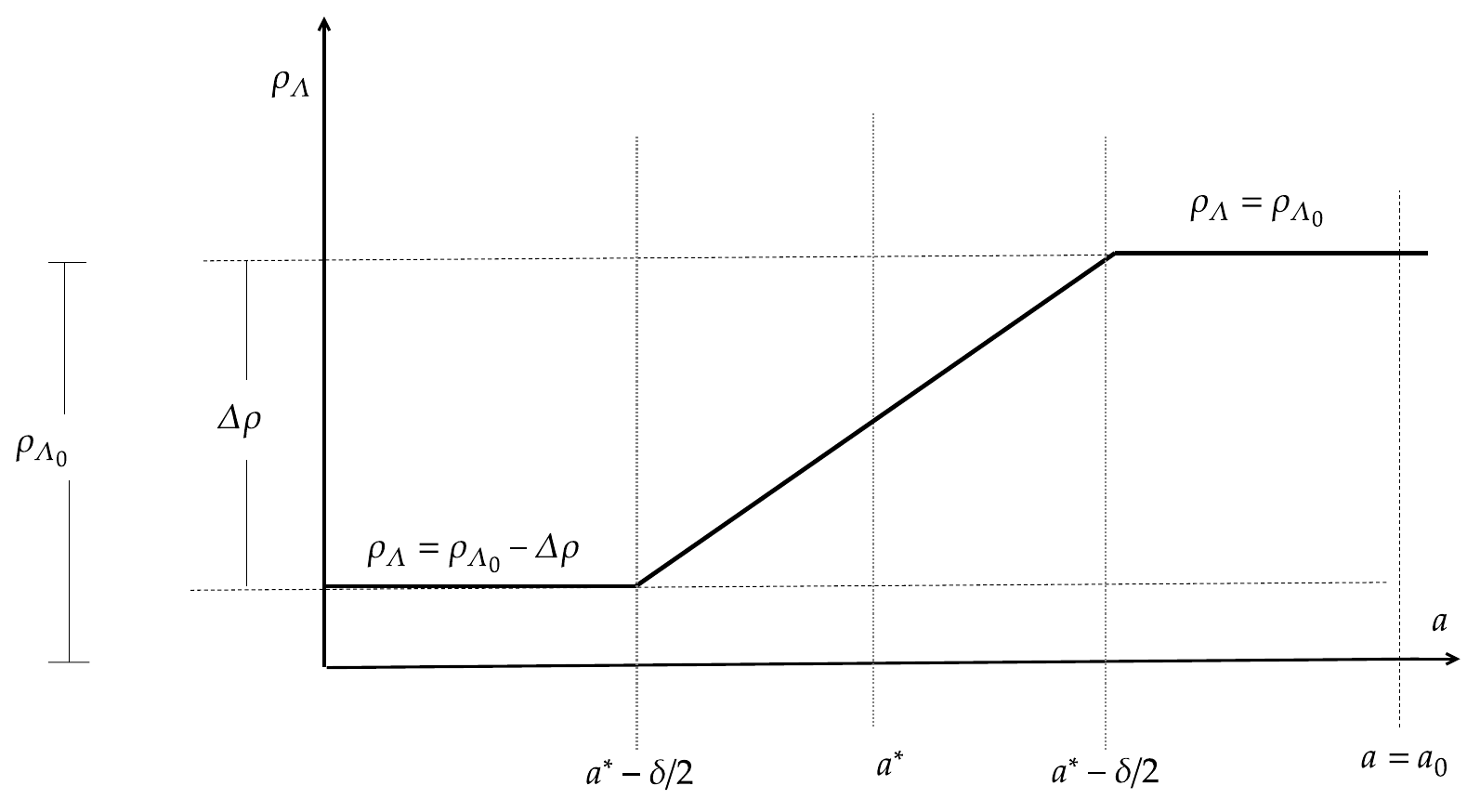}
\caption{Graphical representation of the discrete model, where the parameters $\Delta \rho$, $a^*$, $\delta$ and $\rho_{\Lambda_0}$ define the functional form $\rho_\Lambda$.}
\label{Discrete_model}
\end{figure}

By introducing $\rho_\Lambda = \Lambda/8\pi G$, the continuity equation within the framework of unimodular gravity can be written as
\be
\dot \rho_m  +  3 \frac{\dot a}a \rho_m   = -\dot \rho_{\Lambda}.
\label{eq:rhomatter}
\ee  
The discrete model parametrization (see figure~\ref{Discrete_model}) considers a proposed expression for $\rho_\Lambda$ of the form

\be
	\rho_\Lambda = 
						\begin{cases}
							\quad \rho_{\Lambda_0}-\Delta \rho \quad\quad\quad \quad \quad \quad \quad \quad \quad \quad  a\, \in \,\, (a_{\rm rad},a^*-\delta/2) \,, \\ \\
							\quad  \rho_{\Lambda_0}+ \Delta \rho \left[\dfrac{a- a^* -\delta/2}{\delta} \right] \quad \quad \quad  a \,\in \,\,  (a^* -\delta/2, a^* +\delta/2), \, \\ \\
\quad \rho_{\Lambda_0} \quad \quad \quad\quad \quad\quad \quad \quad \quad \quad \quad \quad  \quad  a \, \in \,\, (a^*+\delta/2,a_0) \,, 
						\end{cases}
\label{DE_1}
\ee
where $a_{\mathrm{rad}}$ and $a_0$ denote the values of the scale factor at the beginning of the radiation-dominated era and at the present time, respectively.

A separation of $\rho_m$ into baryonic and dark matter components is assumed, such that each contributes to the effect in proportion to its cosmic abundance. Following ref.~\cite{Landau:2022mhm}, the baryonic energy density $\rho_b$ is given by

\footnotesize

\be
\rho_b (a)= 
\begin{cases}
\dfrac{\Omega_b \rho_{\rm crit} + \dfrac{\Omega_b}{\Omega_b + \Omega_{c}} \Delta \rho \, \left({a^*}^3  +  \dfrac{a^*\delta^2}{4}\right)}{a^3} \hspace{5cm} a \in \,(a_{\rm rad},a^*-\delta/2), \, \\ \\
 \dfrac{\Omega_b \rho_{\rm crit} + \dfrac{\Omega_b}{\Omega_b + \Omega_{c}} \Delta \rho \, \left({a^*}^3  +  \dfrac{a^*\delta^2}{4}\right)}{a^3}- \dfrac{\Omega_b}{\Omega_b + \Omega_{c}} \dfrac{\Delta\rho}{ 4 \delta}\left[ a  -  \dfrac{(a^*-\delta/2 )^4}{a^3}  \right] a \in \, (a^* -\delta/2, a^* +\delta/2) \,,  \\ \\
\dfrac{\Omega_b \rho_{\rm crit}}{a^3} \hspace{9.3cm} a \in \,\, (a^*+\delta/2,a_0=1).  \quad \,
\end{cases}
\label{omegab}
\ee
\normalsize
Similarly, the cold dark matter energy density $\rho_{\rm cdm}$ takes the form
\footnotesize
\be
\rho_{c}(a) = 
\begin{cases}
\dfrac{\Omega_{c} \rho_{\rm crit} + \dfrac{\Omega_{c}}{\Omega_b + \Omega_{c}} \Delta \rho \, \left({a^*}^3  +  \dfrac{a^*\delta^2}{4}\right)}{a^3} \hspace{5cm} a \in \,\, (a_{\rm rad},a^*-\delta/2), \, \\ \\
\dfrac{\Omega_{c} \rho_{\rm crit} + \dfrac{\Omega_{c}}{\Omega_b + \Omega_{c}} \Delta \rho \, \left({a^*}^3  +  \dfrac{a^*\delta^2}{4}\right)}{a^3}- \dfrac{\Omega_{c}}{\Omega_b + \Omega_{c}}  \dfrac{\Delta\rho}{ 4 \delta} \left[ a  -  \dfrac{(a^*-\delta/2 )^4}{a^3}  \right]  \   a \in \, (a^* -\delta/2, a^* +\delta/2) \,,  \\ \\
\dfrac{\Omega_{c} \rho_{\rm crit}}{a^3} \hspace{9.3cm} a \in \,\, (a^*+\delta/2,a_0=1).  \quad \,
\end{cases}
\label{omegadm}
\ee
\normalsize

\subsection{Continous model}
\label{sec:Continous model}
We now explore a continuous version of the step function. This model is an improvement to the discrete model (figure \ref{Discrete_model}). Following the same procedure, it is possible to solve eq.~\eqref{eq:rhomatter} to obtain an expression for $\rho_m$ by integrating the corresponding differential equation from the beginning of the radiation-dominated era to an arbitrary later time.

Proposing an expression for $\rho_{\Lambda} (a)$ of the form\footnote{It is also possible explore  $ \rho_{\Lambda} (a)  =  \rho_{\Lambda}^0   +\frac{\Delta\rho}{2} \left[\tanh \left(\frac{a-a^*}{\delta}\right)-\tanh\left(\frac{a_0-a^*}{\delta} \right) \right]$ with $\tanh(x)$ growing more rapidly than $\arctan(x)$, the problem is that it is more difficult to provide an analytic expression for equation \eqref{eq:rhomatter} using this function, as it requires the use of the polylogarithm function, which affects drastically computation time.
}
\be\label{DE_2}
 \rho_{\Lambda} (a)  =  \rho_{\Lambda}^0   +\dfrac{\Delta\rho}{\pi} \left[\arctan \left(\dfrac{a-a^*}{\delta}\right)-\arctan\left(\dfrac{a_0-a^*}{\delta} \right) \right],
\ee 
where $\rho_\Lambda^0$ is the current value of the dark energy density.

From eqs.~\eqref{eq:rhomatter} and \eqref{DE_2}, it follows that
\be
\dfrac{d\rho_m}{da}+ \dfrac{3}{a} \rho_m = -\dfrac{\Delta \rho}{\pi \delta} \dfrac{1}{1+\left( \dfrac{a- a^*}{\delta}\right)^2}.
\ee
Solving the previous expression yields

\begin{equation}
\begin{split}
\rho_m = \frac{C}{a^3(t)} 
&- \frac{\Delta \rho}{2\pi} \frac{\delta^3}{a^3(t)} \Bigg\{  \Bigg[
\left(\frac{a(t)}{\delta}\right)
\left(\frac{4a^*+a(t)}{\delta}\right)
-2 \left(\frac{a^*}{\delta}\right)
\left(-3+\left(\frac{a^*}{\delta}\right)^2 \right)
\Bigg] \\
& \times \arctan\!\left( \frac{a^*-a(t)}{\delta} \right) + \left[-1 + 3 \left(\frac{a^*}{\delta}\right)^2 \right]  \log\!\left[1+\left(\frac{a^*-a(t)}{\delta} \right)^2 \right]
\Bigg\}.
\end{split}
\end{equation}
Evaluating at the radiation epoch ($t_{\rm rad}$) determines the integration constant in terms of the energy density at that epoch. Hence, one obtains

\be
C= \rho_m(t_{\rm rad})a^3(t_{\rm rad}) -F(a(t_{\rm rad}), \Delta \rho, \delta,a^*)
\ee
with
\be
\begin{aligned}
F(a(t),\delta,\Delta \rho, &a^*) = -\frac{\Delta \rho}{2\pi}\,\delta^3 \Bigg\{ 
 \Bigg[
\left(\frac{a(t)}{\delta}\right)
\left(\frac{4a^*+a(t)}{\delta}\right)
-2 \left(\frac{a^*}{\delta}\right)
\left(-3+\left(\frac{a^*}{\delta}\right)^2 \right)
\Bigg] \\
& \times \arctan\!\left( \frac{a^*-a(t)}{\delta} \right) + \left[-1 + 3 \left(\frac{a^*}{\delta}\right)^2 \right]
\log\!\left[1+\left(\frac{a^*-a(t)}{\delta} \right)^2 \right]
\Bigg\}.
\end{aligned}
\label{Function_Continous_Model}
\ee
Given that this effect involves an interaction between the total matter density, $\rho_m$, and the dark energy density, $\rho_\Lambda$, constraining the model parameters using CMB data requires separating the matter sector into its baryonic and cold dark matter components. Let us propose the following expressions for the baryon density, $\rho_b(t)$, and the dark matter density, $\rho_c(t)$\footnote{Note that this partition is not unique, but it is motivated, since the second term in eq.~\eqref{Rho1_t}, which contains the function $F(a(t), \delta, \Delta \rho, a^*)$, represents the energy diffusion from the matter component to dark energy. Thus, the parameter $\alpha$ quantifies the fractional contribution of the baryonic and dark matter degrees of freedom involved in this effect.}:
\begin{eqnarray}
    \rho_b(t) &=& \dfrac{\rho_b(t_{\rm rad})a^3(t_{\rm rad})}{a^3(t)} +\dfrac{\alpha }{a^3(t)} \left[F(a(t),\delta,\Delta \rho, a^*)- F(a(t_{\rm rad}), \Delta \rho, \delta,a^*)\right],\label{Rho1_t}\\
    \rho_{c}(t)&=&\dfrac{\rho_{c}(t_{\rm rad})a^3(t_{\rm rad})}{a^3(t)}+\dfrac{(1-\alpha)}{a^3(t)}\left[F(a(t),\delta,\Delta \rho, a^*)- F(a(t_{\rm rad}), \Delta \rho, \delta,a^*)\right],
\label{Rho2_t}    
\end{eqnarray}
where $\alpha$ is a constant satisfying $\alpha < 1$. Defining the present fraction of baryons to dark matter density $\beta=\rho_b(t_0)/\rho_{c}(t_0)$, and imposing equal fraction baryonic energy density and dark matter at radiation domination and at present,

\be
    \beta=\dfrac{\rho_b(t_0)}{\rho_{c}(t_0)}=\dfrac{\rho_b(t_{\rm rad})}{\rho_{c}(t_{\rm rad})}.
\label{beta}
\ee
Hence, from the relation $\rho_b(t_{\rm rad})=\beta \rho_c(t_{\rm rad})$ together with eqs.~\eqref{Rho1_t} and \eqref{Rho2_t}, evaluated at $t=t_{\rm rad}$, one obtains the simple relation

\be
   \beta=\dfrac{\alpha}{1-\alpha}.
\ee
Thus,
\be
    \alpha= \dfrac{\beta}{1+\beta}= \dfrac{\rho_b(t_0)}{\rho_{c}(t_0)+\rho_b(t_0)}=\dfrac{\Omega_b}{\Omega_{c}+\Omega_b}.
\label{alpha}    
\ee
Using eqs.~\eqref{beta} and \eqref{alpha} in eqs.~\eqref{Rho1_t} and \eqref{Rho2_t}, and setting $a_0=1$, one finally obtains

\begin{eqnarray}
    \rho_b(t) &=& \dfrac{\Omega_b \rho_{\rm crit}-\dfrac{\Omega_b}{\Omega_{c}+\Omega_b}F(1,\delta,\Delta \rho, a^*)}{a^3(t)} +\dfrac{\Omega_b}{\Omega_{c}+\Omega_b}\dfrac{1}{a^3(t)} F(a(t),\delta,\Delta \rho, a^*), \label{rhb_C1}\\
    \rho_{c}(t) &=&  \dfrac{\Omega_{c} \rho_{\rm crit}-\left( 1-\dfrac{\Omega_b}{\Omega_{c}+\Omega_b}\right)F(1,\delta,\Delta \rho, a^*)}{a^3(t)} +\dfrac{\Omega_{c}}{\Omega_{c}+\Omega_b}\dfrac{1}{a^3(t)} F(a(t),\delta,\Delta \rho, a^*).  \quad\label{rhdm_C2}
\end{eqnarray}
In general, solving eq.~\eqref{eq:rhomatter} for an arbitrary $\rho_{\Lambda}$ yields the following expression for the total matter energy density
\begin{equation}
    \rho_m= \dfrac{\Omega_b+\Omega_c}{a^3}\rho_{\rm crit}+\dfrac{1}{a^3}\int^{a_0=1}_{a}a^3\dfrac{d\rho_{\Lambda}}{da}da.
\end{equation}
Requiring each component (dark matter and baryonic matter) to contribute proportionally to its cosmic abundance, as in eqs.~\eqref{omegab}, \eqref{omegadm}, \eqref{rhb_C1}, and \eqref{rhdm_C2}. The baryonic and dark matter energy densities are then given by
\begin{align}
        \rho_b&= \dfrac{\Omega_b \rho_{\rm crit}}{a^3} +\dfrac{\Omega_b}{\Omega_c+\Omega_b }\dfrac{1}{a^3}\int^{a_0=1}_{a} a^3 \dfrac{d\rho_{\Lambda}}{da} da,\label{rho_b general}\\
         \rho_c&= \dfrac{\Omega_c \rho_{\rm crit}}{a^3} +\dfrac{\Omega_c}{\Omega_c+\Omega_b }\dfrac{1}{a^3}\int^{a_0=1}_{a} a^3 \dfrac{d\rho_{\Lambda}}{da} da.\label{rho_c general}
\end{align}
These expressions are general and reduce to the cases studied in this work when eqs.~\eqref{DE_1} and \eqref{DE_2} are adopted.

Within the context of UG, a preference for $\Delta\rho_{\Lambda} \neq 0$, indicative of a dynamical dark energy behavior, may be interpreted as a manifestation of violations of local energy conservation in a cosmological context. An important question is how the models compare with standard parametrization schemes and whether they can be encompassed within such schemes. This is discussed in Appendix~\ref{sec: Appendix B}.

\subsection{CMB Temperature Polarization Power Spectrum}
\label{sec:powerspectrum}

\begin{figure}[htbp]
\centering
\includegraphics[width=.63\textwidth]{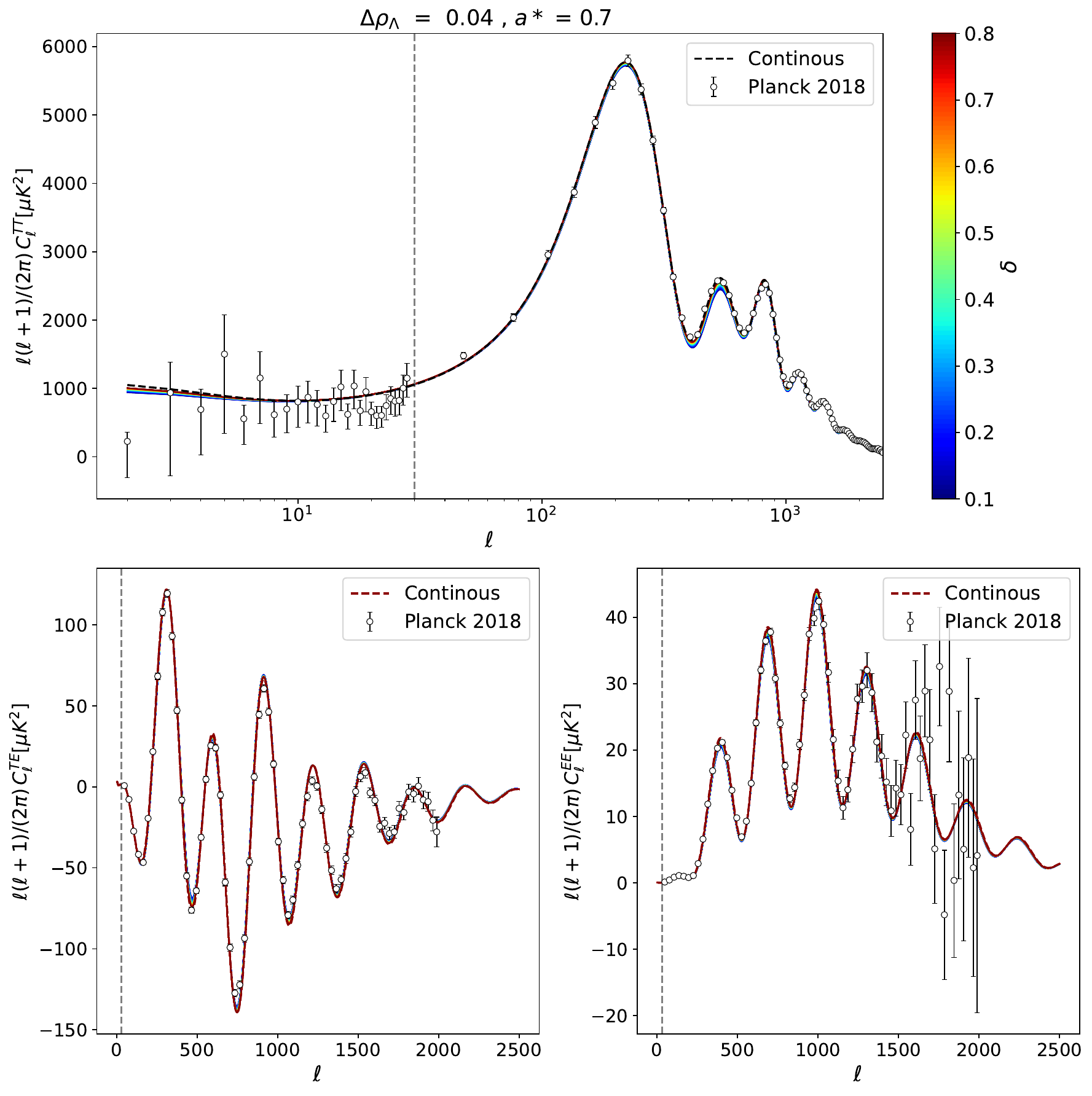}
\caption{Impact of the parameter $\delta$ on the CMB temperature and polarization anisotropy spectra fixing $\Delta \rho_{\Lambda}$ and $a^{*}$. Top: Temperature auto-correlation function ($C_l^{TT}$); Bottom: E-model auto-correlation function ($C_l^{EE}$) and TE correlation function ($C_l^{TE}$).  The vertical gray line separates the high- and low-$\ell \ell$ Planck data. We compare the best-fit results of the continuous model against CMB data.}

\label{delta_impact}
\end{figure}

\begin{figure}
\includegraphics[width=0.49\textwidth]{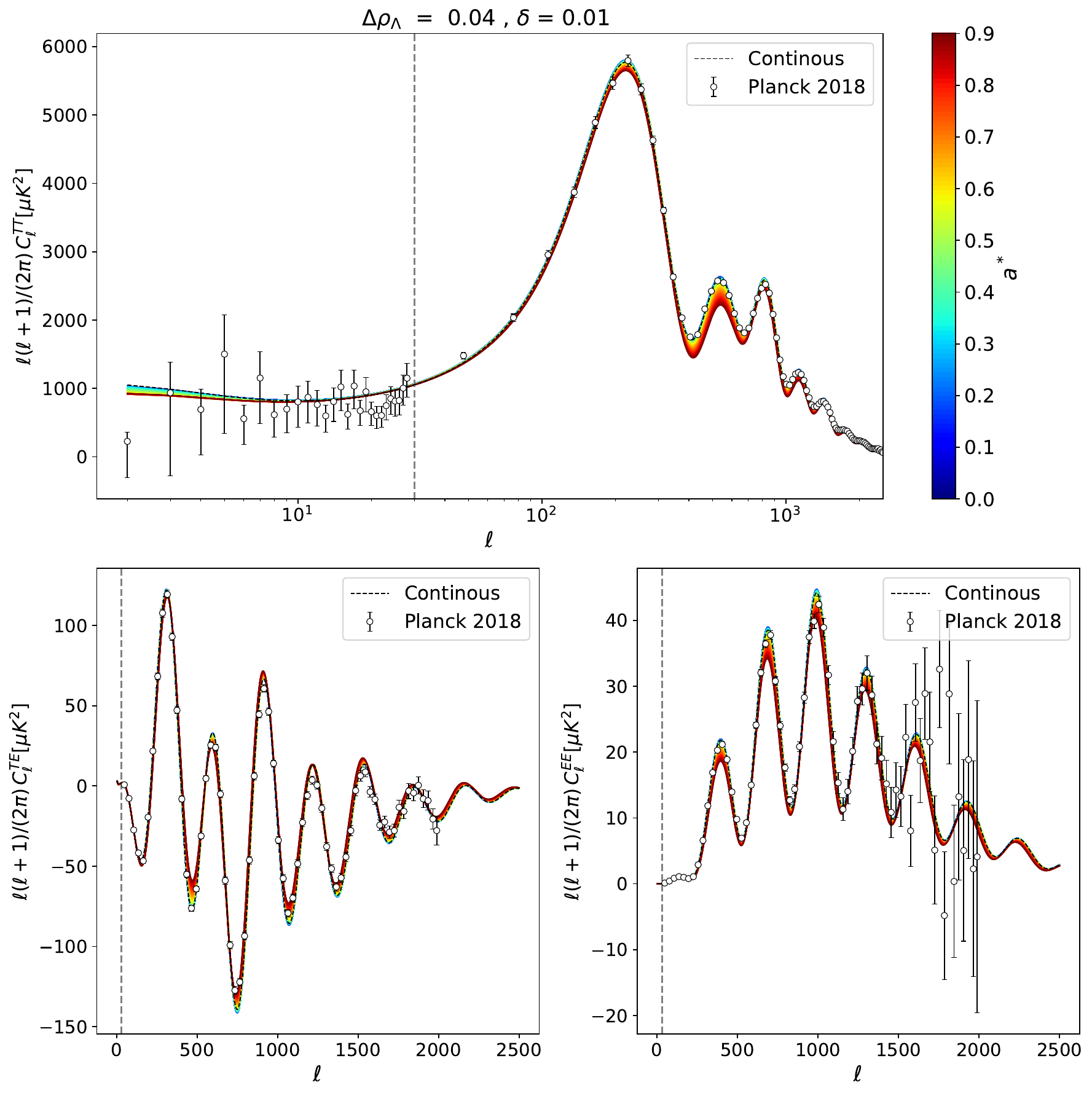}
\includegraphics[width=0.49\textwidth]{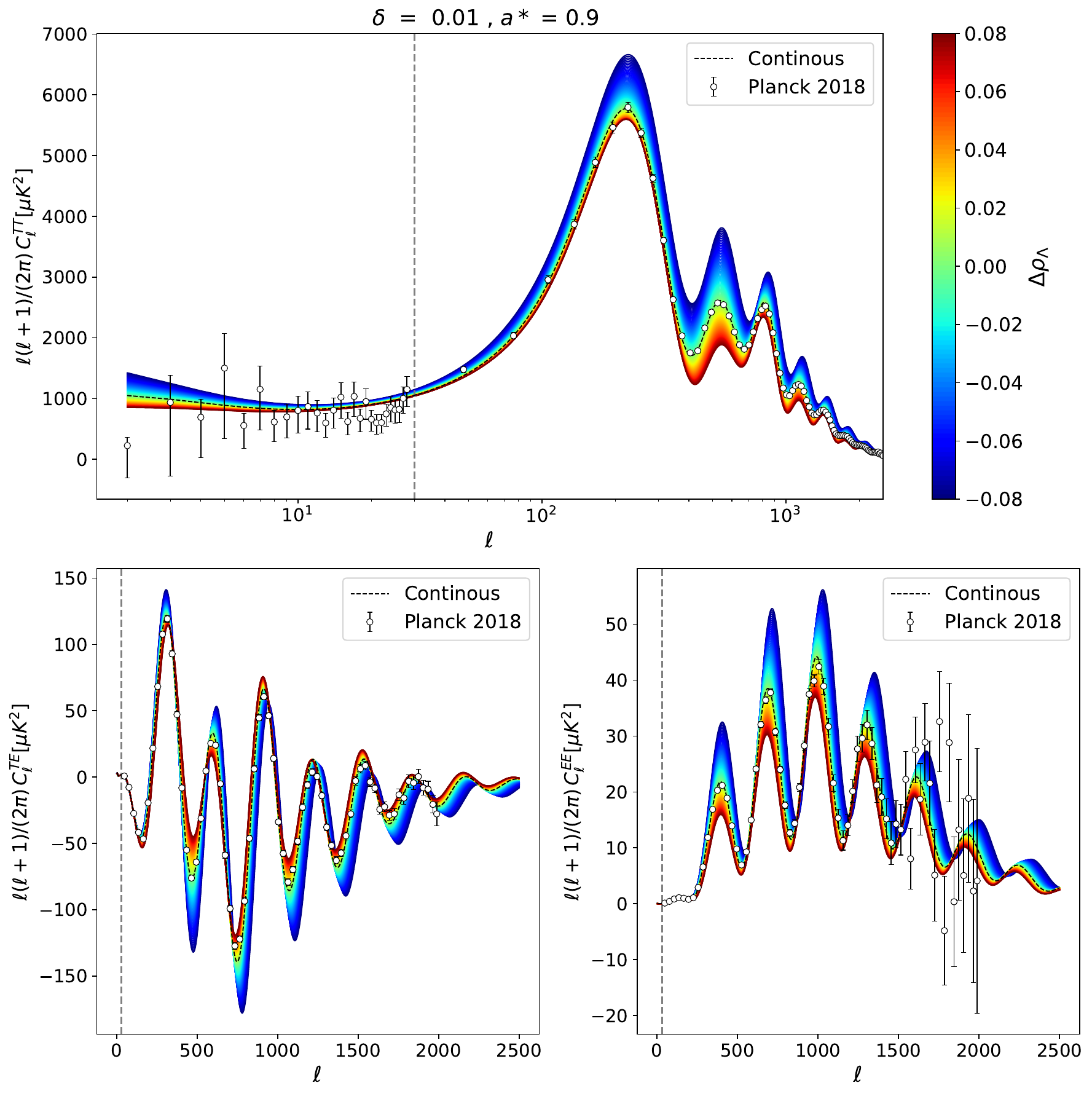}

\caption{Impact of the parameter $a^*$ and $\Delta\rho_{\Lambda}$ in the CMB anisotropy temperature and polarization spectrum. Top: Temperature auto-correlation function ($C_l^{TT}$); Bottom: E-model auto-correlation function ($C_l^{EE}$) and TE correlation function ($C_l^{TE}$).}
\label{a_start_and_Delta_rho_impact}
\end{figure}

The effects of the three parameters, $a^*$, $\delta$, and $\Delta\rho_\Lambda$, on the TT, TE, and EE power spectra are illustrated in figs.~\ref{delta_impact} and \ref{a_start_and_Delta_rho_impact}. As shown, variations in the parameter $\delta$ appear to affect mainly large scales, consistent with the analysis in \cite{Landau:2022mhm}. Meanwhile, the effect of $a^*$ is present at all scales. In addition, the second and third acoustic peaks are also affected by the parameters $a^*$ and $\delta$. At first sight, this may seem surprising, since in the $\Lambda$CDM model their relative amplitudes are mainly determined by the ratio $\Omega_c/\Omega_b$. However, if we look closely, for example, in the matter and baryon fractions in the discrete model, at the beginning of the universe (CMB physics), they are given by

\begin{eqnarray}
\Omega_b^{\rm early}&=& \Omega_b  + \dfrac{\Omega_b}{\Omega_b + \Omega_{c}}\frac{\Delta \rho}{\rho_{\rm crit}} \left({a^*}^3  +  \dfrac{a^*\delta^2}{4}\right), \label{Discrete_Omega_b_early}
 \\
\Omega_{c}^{\rm early}&=& \Omega_{c}  + \dfrac{\Omega_{c}}{\Omega_b + \Omega_{c}} \frac{\Delta \rho}{\rho_{\rm crit}} \left({a^*}^3  +  \dfrac{a^*\delta^2}{4}\right).
\label{Discrete_Omega_cdm_early}
\end{eqnarray}
And for the continuous model 

\begin{eqnarray}
    \Omega_b^{\rm early} &\approx& \Omega_b +\dfrac{1}{\rho_{\rm crit}}\dfrac{\Omega_b}{\Omega_{c}+\Omega_b} [F(-\infty,\delta,\Delta \rho, a^*)-F(1,\delta,\Delta \rho, a^*)], \label{Continous_Omega_b_early}
    \\
            \Omega_{c}^{\rm early} &\approx&  \Omega_{c} -\dfrac{1}{\rho_{\rm crit}}\left( 1-\dfrac{\Omega_b}{\Omega_{c}+\Omega_b}\right)F(1,\delta,\Delta \rho, a^*) +\dfrac{1}{\rho_{\rm crit}}\dfrac{\Omega_{c}}{\Omega_{c}+\Omega_b} F(-\infty,\delta,\Delta \rho, a^*). \quad \label{Continous_Omega_cdm_early}
\end{eqnarray}

Both definitions are viable, since the dark matter fraction is expected to remain nearly constant at early times, $\rho_{b,c}=\Omega^{\rm early}_{b,c}\rho_{\rm crit}/a^3$, until the energy transferred from black holes' spin becomes significant. In the discrete case, for $a^*+\delta/2<1$, $\rho_{\Lambda}$ reaches a constant value before the present epoch, and dark matter and baryonic matter subsequently evolve according to $\rho_{b,c}=\Omega_{b,c}\rho_{\rm crit}/a^3$. For the continuous case, the situation is more subtle. Nevertheless, this behavior remains valid provided that the parameters $\delta$ and $a^*$ yield a continuous dark energy evolution that approximates this asymptotic behavior, such that $d\rho_{\Lambda}/da \approx 0$ in the vicinity of both $a=0$ and $a=1$ in eqs.~\eqref{rho_b general} and \eqref{rho_c general}.

Hence, unimodular parameters are linked to the representative $\Omega_{b}^{\rm early}$, $\Omega_{c}^{\rm early}$, which are the representatives of  $\Omega_{b}$, $\Omega_{c}$ in the late universe. Since $\Omega_{b}^{\text{early}}$ and $\Omega_{c}^{\text{early}}$ are determined by the unimodular parameters, their imprint on the CMB can be interpreted as an effective combination of $\Omega_{b}$ and $\Omega_{c}$ within the standard $\Lambda$CDM framework, resulting in a modification of the acoustic peak amplitudes. The effect modifies the physics of the primordial plasma and impacts the distance measures associated with both the early and late-time evolution of the Universe. The impact of the unimodular parameters in the CMB temperature (TT) and polarization (EE, TE) power spectrum can be understood by drawing insight from the discrete counterpart of the continuous model. The parameters $\delta$ and $a^{*}$, which control the duration of the transition, according to figures~\ref{delta_impact} and \ref{a_start_and_Delta_rho_impact} play a subdominant role. This behavior is also evident from the discrete expressions in eqs.~\eqref{Discrete_Omega_cdm_early} and \eqref{Discrete_Omega_b_early}, where the contribution scales linearly with $\Delta\rho$, cubically with $a^{*}$, and quadratically with $\delta$.

\subsection{ Data and analysis choices}
In this analysis, we use measurements of the Cosmic Microwave Background, temperature anisotropies, and polarization through the Planck (2018) and lensing reconstruction power spectrum with the `click` and `plik` likelihoods \cite{Planck:2019nip, Planck:2018lbu}, the Baryon Acoustic Oscillation measurements from Data Release 2 (DR2) of the Dark Energy Spectroscopic Instrument (DESI) \cite{DESI:2025zgx}, together with supernova distance measurements from the five-year observation of the Dark Energy Survey (DES Y5) (for the individual set analysis, see Appendix~\ref{sec:Appendix A}).

We vary the three parameters of the unimodular model ($a^*$, $\delta$, $\Delta \rho_\Lambda$), which correspond to the onset, duration, and strength of the effect\footnote{With $\Delta \rho_\Lambda \equiv \dfrac{8\pi G}{3} \dfrac{\Delta \rho}{100^2} = \dfrac{\Delta \rho h^2}{\rho_{\rm crit}}$.} described in Sections~\ref{sec:Discrete model} and~\ref{sec:Continous model}. We also vary usual cosmological parameters; the baryon density, $\Omega_b h^2$ the cold dark matter density, $\Omega_{c}$, the angular sound horizon distance, $\theta_{\star}$, the optical depth, $\tau$, the primordial spectral index, $n_s$, and the scalar primordial amplitude fluctuation $A_s$, assuming broad flat priors given by the table \ref{tab:prior}. We use
Metropolis MCMC sampler implemented in Cobaya \cite{Torrado:2020dgo}, employing the fast–slow parameter hierarchy algorithm \cite{Lewis:2013hha} and plotted using GetDist \cite{2019arXiv191013970L}. To explore the possibility of a diffusion mechanism from dark matter to dark energy consistent with the black hole diffusion scenario, we also consider the case $\Delta \rho_\Lambda \geq 0$, adopting the prior $\Delta \rho_\Lambda \in [0,0.2]$.

\begin{table}[htbp]
\centering
\begin{tabular}{c c c}
\hline
\hline
Parameter
& Discrete Model %
& Continuous Model  %
\\
\hline
\hline
$\Omega_{b}h^2$
& $\mathcal{U}(0.005,1)$ 
& $\mathcal{U}(0.005,1)$  %
\\ 
$\Omega_{c}h^2$
& $\mathcal{U}(0.001,0.99)$ %
& $\mathcal{U}(0.001,0.99)$ %
\\
$\tau$
& $\mathcal{U}(0.01,0.8)$ %
& $\mathcal{U}(0.01,0.8)$ %
\\
$\ln(10^{10}A_s)$
& $\mathcal{U}(1.61,3.91)$ %
& $\mathcal{U}(1.61,3.91)$ %
\\
$n_s$
& $\mathcal{U}(0.8,1.2)$ %
& $\mathcal{U}(0.8,1.2)$ %
\\
$100\times\theta_s$
& $\mathcal{U}(0.5,10)$ %
& $\mathcal{U}(0.5,10)$ %
\\
$\Delta\rho_{\Lambda}$
& $\mathcal{U}(-0.2,0.2)$ %
& $\mathcal{U}(-0.2,0.2)$, $\mathcal{U}(0,0.2)$ %
\\
$a^*$
& $\mathcal{U}(0.02,1)$ %
& $\mathcal{U}(0.02,1)$ %
\\
$\delta$
& $\mathcal{U}(0.02,1)$  %
& $\mathcal{U}(0.002,1)$ %
\\
$\Omega_k$
& $-$ %
& $\mathcal{U}(-0.3,0.3)$ %
\\
\hline
\hline
\end{tabular}
    \caption{Priors on cosmological and unimodular parameters.  We assume large flat priors.  In the continuous model, the prior domain was changed to $\delta \in [0.002, 1]$. As mentioned, this is because the parameter $\delta$ in the two models is not directly related, and no unique mapping exists between them. In fact, a value of $\delta = 0.02$ in the continuous model corresponds to a smooth transition spread over a large domain, which makes it necessary to modify the domain of exploration.} 
    \label{tab:prior}
\end{table}

\vspace{0.3cm}

\section{Results }
\label{sec:Results}

\subsection{Cosmological Constraints on diffusion models in a flat universe}
\label{sec:Discrete and Continuous parameter model constraint}
\begin{figure}
\centering
\includegraphics[width=0.95\textwidth]{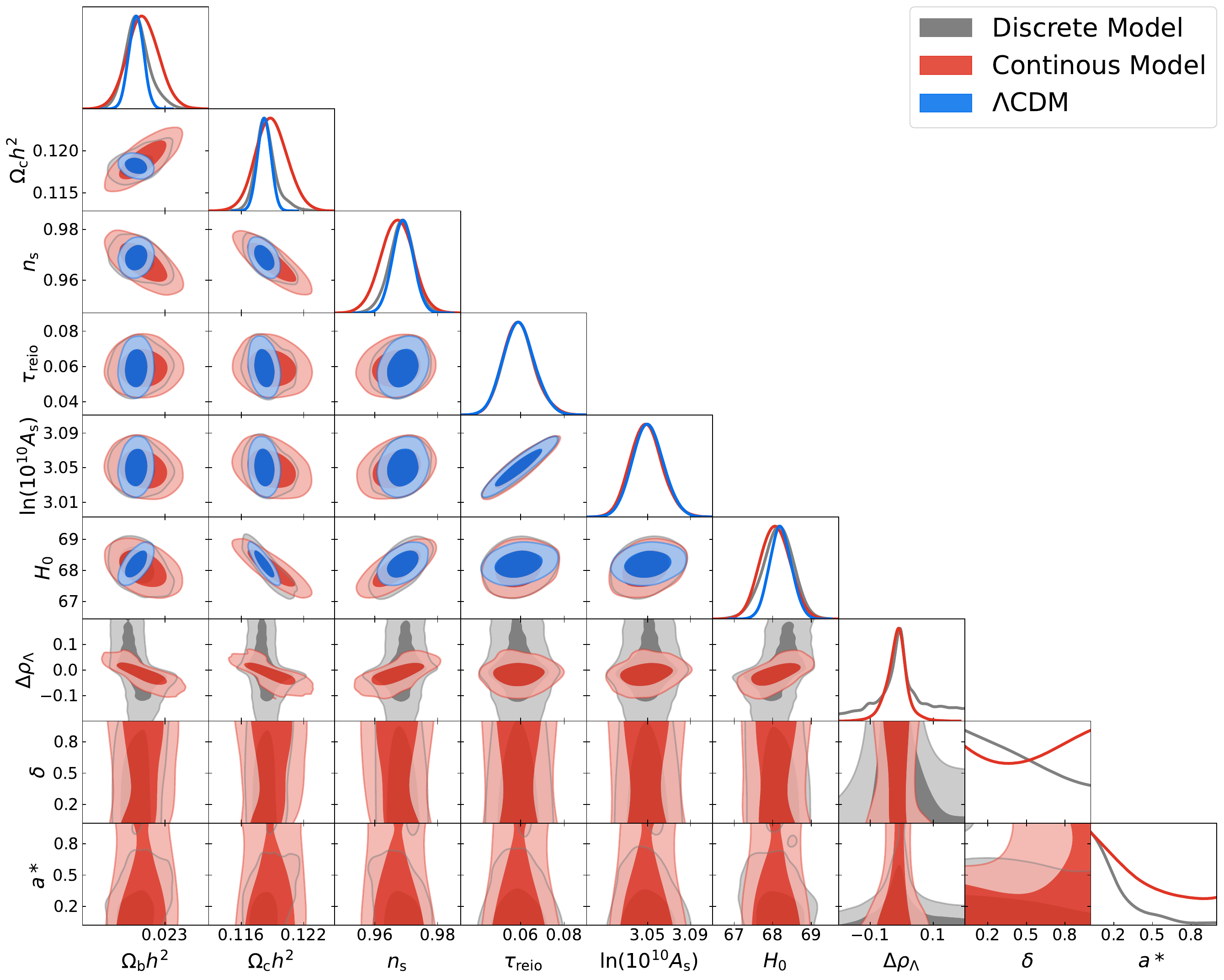}  
\caption{Constraints on the cosmological and unimodular parameters for discrete (gray) and continuous models (red). The blue region represents confidence intervals obtained using the $\rm\Lambda CDM$ model. The filled strong and light blue region shows the $68\%$ and $95\allowbreak\%$ confidence intervals.}
\label{Complete_No_Curvature_fig1}
\end{figure}

In figure \ref{Complete_No_Curvature_fig1}, we plot the confidence contours 
for the cosmological parameters obtained for the continuous and discrete unimodular models. For reference, we include the cosmological constraints for the $\rm\Lambda CDM$. The specific parameters for the models based on unimodular gravity are also plotted, showing consistency between the two models. However, care must be taken when directly relating the parameter $\delta$ appearing in both models, as the parameter spaces explored are not equivalent, making a direct comparison problematic. In the discrete model, this parameter is associated with the period during which diffusion occurs, characterized by the interval $(a^* - \delta/2,\, a^* + \delta/2)$. In the continuous case, however, such an interval can only be identified approximately from the asymptotic behavior of the dark energy density function. This is reflected in the flat prior domain chosen in both cases for the MCMC analysis (see table \ref{tab:prior}). In the case of an abrupt transition, $\delta = 0.02$ in the discrete model, smaller values are required in the continuous model to reproduce such a localized transition. In particular, for small values of $\delta$, $\delta_{\rm continuous} \approx \delta_{\rm discrete}/10$ appears to be sufficient to reproduce the discrete behavior at a graphical level.

The posteriors show a good agreement with $\rm\Lambda CDM$ model in the standard cosmological parameters $\Omega_bh^2, \Omega_{c} h^2, \tau_{\rm reio}, H_0$ and $n_s$, which is expected given that according to the constraints only a small fraction of the total matter is diffused into the dark energy sector, and hence $\Delta \rho_{\Lambda} \ll \Omega_{b,c} h^2$. For the discrete model, the mean values indicate an early transition starting at $z \approx 165.7$ ($a\approx 0.006$) and ending at $z \approx 1.27$ ($a\approx0.44$), with $\Delta \rho_{\Lambda} \approx 0.001$. Previous analyses reported best-fit parameter values corresponding to a transition occurring at $z \approx 0.45-0.10$ corresponding to $a^* = 0.8$ and $\delta = 0.22$, and $68\%$ confidence levels of $\delta<0.65$ and $a^*>0.4$. Meanwhile, the data show a preference for a diffusion effect with a midpoint around $z \approx 1.56$ ($a^* \approx 0.39$) and $\delta \approx 0.52$ for the continuous model, which on average appears to favor $\Delta \rho_{\Lambda} < 0$. In both cases, the diffusion starts, or becomes relevant, between the epoch of last scattering and late-time observables. We present in table \ref{tab:Comparison_Model_table} the mean, $1\sigma$ uncertainties, and the best fit values for both discrete and continuous models, comparing them with
those of the standard $\rm \Lambda CDM$ model. In particular, the negative mean value for $\Delta \rho_{\Lambda}$ in the case of a flat Universe in the continuous model is contrary to the expected positive value corresponding to a diffusion from matter sector to dark energy sector. This indicates a degeneracy, requiring a higher matter density at late times than that inferred from CMB data.

Although a negative $\Delta\rho_{\Lambda}$ may appear problematic, as mentioned, black holes may not be the only source of the cosmological constant. It may also be associated with processes underlying spontaneous matter localization, as in objective collapse theories of quantum mechanics, such as the CSL theory. Other alternatives include particle motion affected by granular friction or, more generally, physical mechanisms that lead to violations of energy--momentum conservation. In such cases, the effective non-conservation at the theoretical level would be given by
\begin{equation}
    \mathbf{J} = \mathbf{J}_{\rm CSL} + \mathbf{J}_{\rm particle} + \mathbf{J}_{\rm BH} + \mathbf{J}_{\rm other}.
\label{eq. current}
\end{equation}
In fact, according to \cite{2017arXiv171105183P, Josset:2016vrq}, some estimates for kinetic particle diffusion yield
$\mathbf{J}_{\rm particle} \approx 2\pi \alpha \frac{T}{m_p^2} R^2 dt$ and $\mathbf{J}_{\rm CSL} = -8\pi G\xi_{\mathrm{CSL}} \rho_b \, dt$,
where $T$ is the temperature, $R$ is the Ricci scalar, and $\xi_{\rm CSL}$ is a positive constant. So that $\mathbf{J}_{\rm CSL}$ contributes to decreasing the cosmological constant and consequently $\rho_\Lambda$.

\begin{table}[htbp]
\centering
\caption{Analysis constraints for the model parameters using the data set Planck (2018) + lensing + DESI DR2 + DES Y5. We also compare the results with the standard $\Lambda$CDM cosmological model. Upper/lower limits correspond to $68\%$ C.L. intervals.}
\label{tab:Comparison_Model_table}

\scalebox{0.7}{
\begin{tabular}{|c|c|c|c|c|c|c|}
\hline
\multicolumn{1}{|c|}{Parameter} &
\multicolumn{2}{c|}{$\Lambda$CDM} &
\multicolumn{2}{c|}{Continuous Model} &
\multicolumn{2}{c|}{Discrete Model} \\
\hline
{ }&
{mean value and }&
{bestfit}&
{mean value and }&
{bestfit}&
{mean value and}&
{bestfit}
\\
 & $68 \%$ confidence levels & & $68 \%$ confidence levels & &  $68 \%$ confidence levels & \\
\hline
$100\,\Omega_b h^2$
& $2.2508 \pm 0.0125$
& $2.2475$
& $2.2626 \pm 0.0271$
& $2.2715$
& $2.2532 \pm 0.0218$
& $2.2527$
\\
$\Omega_{\rm c} h^2$
& $0.1182 \pm 0.0006$
& $0.1183$
& $0.1189 \pm 0.0015$
& $0.1188$
& $0.1183 \pm 0.0010$
& $0.1178$
\\ 
$\tau$
& $0.0592 \pm 0.0072$
& $0.0520$
& $0.0592 \pm 0.0072$
& $0.0622$
& $0.0591 \pm 0.0070$
& $0.0610$
\\
$\ln(10^{10} A_s)$
& $3.0504 \pm 0.0144$
& $3.0372$
& $3.0488 \pm 0.0149$
& $3.0554$
& $3.0499 \pm 0.0144$
& $3.0538$
\\
$n_s$
& $0.9689 \pm 0.0033$
& $0.9676$
& $0.9669 \pm 0.0051$
& $0.9669$
& $0.9684 \pm 0.0039$
& $0.9707$
\\
$\Delta \rho_{\Lambda}$
& $-$ 
& $-$ 
& $-0.0159 \pm 0.0349$
& $-0.0151$
& $0.0001 \pm0.084$
& $-0.0078$
\\
$a^*$
& $-$ 
& $-$ 
& $0.3933 \pm 0.2790$
& $0.9935$
& $0.2232 \pm 0.1920$
& $0.0352$
\\
$\delta$
& $-$ 
& $-$ 
& $0.5266 \pm 0.3007$
& $0.8819$
& $0.4352\pm 0.2736$
& $0.6638$
\\
$\Omega_k$
& $-$ 
& $-$ 
& $-$ 
& $-$ 
& $-$ 
& $-$ 
\\
$H_0$ [km\,s$^{-1}$\,Mpc$^{-1}$]
& $68.2033 \pm 0.2845$
& $68.0979$
& $68.0616 \pm 0.3897$
& $68.1404$
& $68.1348 \pm 0.4032$
& $68.2244$
\\
\hline
\end{tabular}
}
\end{table}

\subsection{Cosmological Constraints on diffusion models with curvature}
\label{sec:Curvature Extension Model}

The results of a negative value $\Delta\rho_{\Lambda}$ considering a flat Universe and a continuous transition indicate a need for more matter at late time in order to account for the observations. In particular, the discrepancy between the density matter needed for accounting for the temperature power spectrum at last scattering surface and the one needed to account for the lensing observed in the polarization power spectra \cite{DiValentino2020}.
In recent analysis, curvature has been shown to alleviate the tension between the CMB and DESI BAO data, being particularly effective in alleviating the tension of the sum of neutrino masses $(\sum m_\nu)$  with the particle physics oscillation experiments ($m\sim0.6\ \mathrm{eV}$) constraint \cite{Chen:2025mlf}, allowing for larger values of $\Omega_m$, thereby enhancing more clustering and then permitting positive neutrino mass. Also, studies over the reionization scenario, integrating Pop III stars \cite{Tan2025}, can change the value of $\Omega_m$ required to reproduce the CMB temperature power spectrum. In a more realistic situation, it is natural to extend our model to the case where the spatial curvature $\Omega_k \neq 0$.

\begin{figure}
\centering
\includegraphics[width=0.95\textwidth]{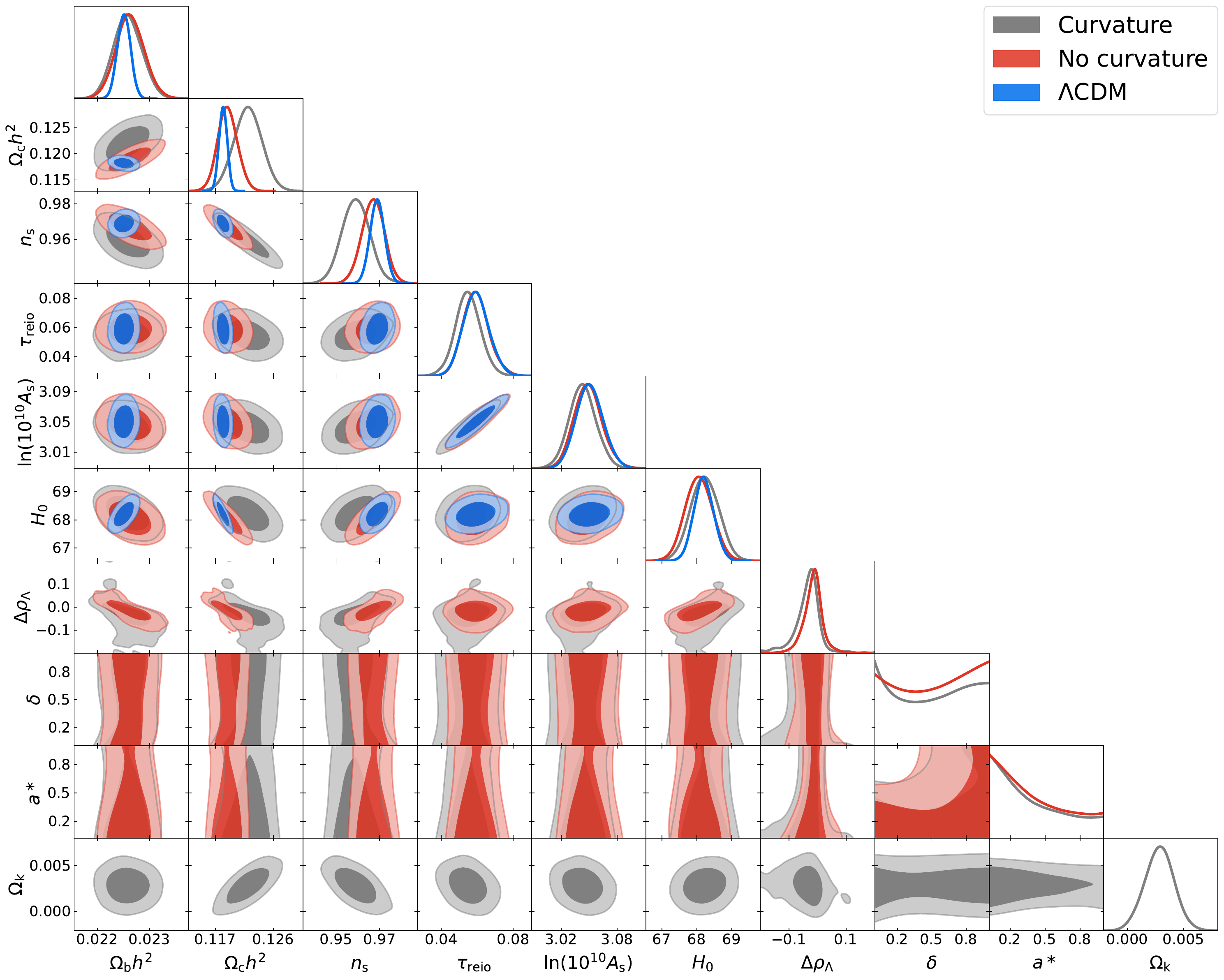}  
\caption{Posterior results of the cosmological parameter and unimodular parameter for the continuous model with curvature (Grey) and without curvature (Red). $\rm\Lambda CDM$ standard model is also plotted for comparison.}
\label{fig:curvature1}
\end{figure}

\begin{table}[htbp]
\centering
\caption{Constraints from the analysis, including spatial curvature for the continuous model parametrization.}
\label{tab:curvature}

\scalebox{0.89}{
\begin{tabular}{|c|c|c|c|c|c|}
\hline
Parameter&
{mean value and }&
{bestfit}&
Parameter&
{mean value and }&
{bestfit}\\
 & $68 \%$ confidence levels &  & & $68 \%$ confidence levels &  \\
\hline
$100\,\Omega_b h^2$
& $2.2576 \pm 0.0271$
& $2.2326$
&$\Delta \rho_{\Lambda}$
& $-0.0339 \pm 0.0473$
& $-0.0063$
\\
$\Omega_{\rm c} h^2$
& $0.1220 \pm 0.0021$
& $0.1209$
& $a^*$
& $0.3821 \pm 0.2783$
& $0.4532$
\\ 
$\tau$
& $0.0548 \pm 0.0071$
& $0.0525$
& $\delta$
& $0.5056 \pm 0.3063$
& $0.7690$
\\
$\ln(10^{10} A_s)$
& $3.0429 \pm 0.0145$
& $3.0383$
& $\Omega_k$
& $0.0028 \pm 0.0013$
& $0.0032$
\\
$n_s$
& $0.9591 \pm 0.0064$
& $0.9611$
& $H_0$ [km\,s$^{-1}$\,Mpc$^{-1}$]
& $68.2281 \pm 0.4042$
& $68.6444$
\\
\hline
\end{tabular}
}
\end{table}
An extended analysis, varying curvature, is developed using the same data set as for the continuous model, and the posterior plots are shown in figure~\ref{fig:curvature1}. In table \ref{tab:curvature}, we present the best-fit and mean parameter values in the model with curvature.
When we compare these results with those obtained in Section~\ref{sec:Discrete and Continuous parameter model constraint}, we find that the mean values correspond to a transition midpoint at $a \approx 0.38$ ($z \approx 1.63$) with $\Delta\rho_{\Lambda}\approx-0.03$. We also find a mean value of $\Omega_k \simeq 2.8 \times 10^{-3}$, indicating a preference for negative spatial curvature.

\begin{figure}[htbp]
\centering
\includegraphics[width=0.95\textwidth]{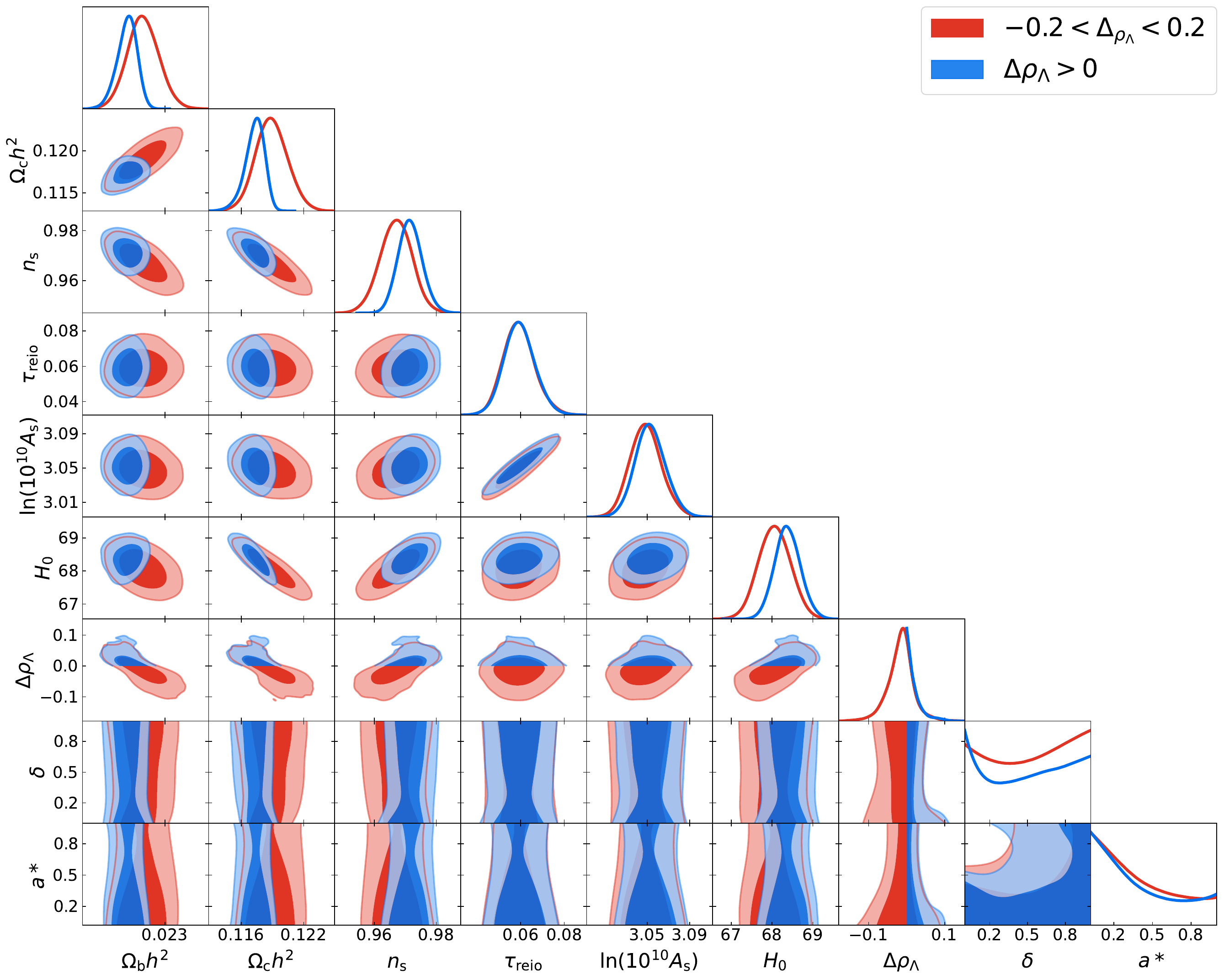}

\caption{Posterior results with a flat prior $\Delta \rho_{\Lambda} \geq 0$, compared with the standard case of Section A $(-0.2 \leq \Delta \rho_{\Lambda} \allowbreak \leq 0.2)$.}
\label{DeltaRhoPositive}
\end{figure}

Similarly, to explore the point density distribution for positive values of $\Delta\rho_\Lambda$, we run an analysis with the constraint $\Delta\rho_\Lambda \geq 0$. As shown in figure~\ref{DeltaRhoPositive}, no significant differences are observed in the standard cosmological parameters. This result is important since parameters such as $\Omega_b$ are strongly constrained by Big Bang Nucleosynthesis (BBN). More precisely, $\Omega_{b}^{\mathrm{early}}$ is the quantity that BBN is expected to constrain. However, because the effects of the model at early times do not significantly modify the matter fractions, no tension with BBN is expected\footnote{For example, for the discrete model we obtain mean values $100\,\Omega_{b}^{\rm early}=0.4855$ and $\Omega_{c}^{\rm early}=0.2551$. For $h=0.6813$, this corresponds to $100\omega_{b}^{\rm early}\equiv100 \Omega_{b}^{\rm early}h^2=2.254$ and $\omega_{c}^{\rm early}\equiv \Omega_{c}^{\rm early}h^2=0.1184$.}. However, for $\Delta \rho_{\Lambda} \geq 0$, the posterior of the $H_0$ parameter slight shifts from $68.06 \pm 0.39$ to $68.38 \pm 0.32$.

\subsection{Model Comparison}
To compare the UG model balancing goodness of fit and the complexity, three statistical information criteria were used: the Akaike Information Criterion (AIC) \cite{1100705}, the Bayesian Information Criterion (BIC) \cite{doi:10.1177/0049124104268644}, and the Deviance Information Criterion (DIC) \cite{Liddle:2007fy}. The Akaike Information Criterion (AIC) is defined as $\text{AIC}=\chi_{\rm min}^2 +2k$, where $\chi_{\rm min}^{2}$ denotes the minimum chi-squared value, and $k$ is the number of parameters in the model. The additional term penalizes model complexity to avoid overfitting. The DIC is defined as: $\text{DIC}=2\bar{D}-\hat{D}$, where $\hat{D}= -2\ln(\mathcal{L}_{\rm max})$ is the deviance evaluated at the maximum likelihood and $\bar{D}=-2\langle\ln(\mathcal{L}_{\rm max})\rangle$ is the mean deviance averaged over the posterior distribution of the
parameters. BIC criterion is built considering not only the complexity but also the data, it uses $\chi^2_{\rm min}+k\ \ln(n)$,  where $n$ is the total number of data points. We compare the fits with the reference $\rm\Lambda CDM$ model by evaluating the differences in the information criterion and using the commonly used Burnham \& Anderson scale \cite{burnham2003model}.

We present the results of the model comparison estimation in the table~\ref{tab: Information Criteria}. The $\Delta\chi^2$ exhibits an expected improvement in the best-fit likelihood, adding more complexity in the models. We can note that the BIC is strongly in disfavour of the UG models. 
However, its complexity penalty scales as $k\ln n$, and the interpretation of n becomes less straightforward for highly correlated datasets such as the CMB power spectra. Consequently, the BIC may provide an overly conservative assessment of model complexity in this context \cite{schwarz1978}.
We therefore also consider the DIC, which characterizes model complexity through the posterior distribution rather than through the nominal number of parameters \cite{Liddle2007}. The DIC indicates a slight-to-moderate preference for the UG models, particularly when curvature is included. However, DIC is known to be most reliable for approximately Gaussian, unimodal posteriors with limited prior sensitivity. Inspection of the posterior distributions shows that these conditions are only partially satisfied. Consequently, the DIC results should be interpreted cautiously.
Given the limitations associated with both the BIC and DIC in the present analysis, we consider the AIC \cite{akaike1974} to provide the most balanced assessment of the trade-off between goodness-of-fit and model complexity. The current data do not provide strong evidence in favor of UG over $\mathrm{\Lambda CDM}$. Nevertheless, UG achieves a comparable fit quality while offering a modest reduction of the $H_0$ tension, especially when considering a curvature.

\begin{table}[htbp]
\label{tab: Information Criteria}
\centering
\begin{tabular}{|c|c|c|c|c|}
\hline
\multicolumn{1}{|c|}{} &
\multicolumn{1}{c}{} &
\multicolumn{2}{|c|}{No curvature} &
\multicolumn{1}{|c|}{Curvature} \\
\hline
Parameter
&$\Lambda$CDM   %
& Continuous Model %
& Discrete Model %
& Continuous Model  %
\\
\hline
$\Delta\chi^2$
& $0$ 
& $-0.92$ %
& $-2.0736$ %
& $-6.5756$ %
\\
$\mathrm{\Delta BIC}$
& $-$ %
& $26.2990$ %
& $25.1454$ %
& $29.7165$ %
\\
$\mathrm{\Delta DIC}$
& $-$ %
& $-1.5949$ %
& $-0.7888$ %
& $-6.3180$ %
\\
$\mathrm{\Delta AIC}$
& $-$ %
& $5.0799$ %
& $3.9263$ %
& $1.4243$ %
\\
\hline
\end{tabular}
\caption{Information criteria for continuous and discrete models, with and without spatial curvature. Here $\Delta Q = Q_{\mathrm{ UG\  model}} - Q_{\rm\Lambda CDM}$, for $Q=\{\chi^2, \mathrm{AIC}, \mathrm{BIC}, \mathrm{DIC}\}$.}
\end{table}

\section{Discussion}
\label{sec:Discussion}

Given that we consider diffusion as an effective manifestation of interactions with black-granularity, our current understanding of black hole channel formation and evolution could provide valuable insights into the validity of the model. The effect of the diffusion mechanism is expected to become more significant once a sufficiently large black hole density ($\rho_{\rm BH}$) is present. Black hole formation can occur through three primary channels:
(i) In the early universe, by complex cooling mechanisms involving $H_2$ gas, stars can form and subsequently collapse. In principle, these may be detected as Population III star remnants, with low-metallicity stars collapsing on timescales of approximately $t \sim 2$ Myr;
(ii) the collapse of disk gas once matter dominates over radiation, leading to the formation of structures that eventually produce black holes; and (iii) originating from the initial conditions at the early universe through the collapse of primordial density perturbations, giving rise to primordial black holes (PBHs). Empirically evidence of such PBH detected as Quasar at redshifts $z \lessapprox 10$ \cite{SDSS:2001emm, 2003AJ....125.1649F, 2011Natur.474..616M, 2021ApJ...907L...1W, Castellano_2023, Bogdan:2023ilu}. Although PBHs formed through the collapse of primordial perturbations are expected to be generated with negligible spin, in which case a diffusion mechanism like that associated with the interaction of BH-granularity becomes negligible. However, it is well known that interactions between spinless black hole pairs can produce a final state with significant spin. In such a situation, a diffusion mechanism that begins at early times after the Big Bang, as obtained in our analysis, does not seem so unusual. Alternative scenarios indicate that PBHs may also form through bubble collisions during first-order phase transitions in old inflationary models, as well as via the collapse of domain walls in more modern inflationary frameworks \cite{Carr:2020gox}. 

An important part of the analysis is to investigate results consistent with the black hole friction mechanism. These values are associated with two requirements. First, the diffusion process is assumed to transfer energy from the matter sector to the dark energy sector. Second, an asymptotic recovery of the standard $\Lambda$CDM behavior in the limit $a\rightarrow 0$ is desired, consistent with a scenario in which black holes form in the early Universe—either through the collapse of primordial density perturbations or via some of the mechanisms discussed above—require time before contributing significantly to the dark energy density. Black holes formed from primordial density perturbations must subsequently grow through accretion and mergers and acquire angular momentum. The same applies to those formed through stellar or matter collapse, which only become significant contributors after their population has sufficiently developed and enough spin energy has been diffused. Eventually, black holes are expected to gradually lose their rotational energy through the mechanism described by eq.~\eqref{Spin}, until no further energy can be extracted. As the black hole formation rate decreases and the available spin-energy reservoir is depleted, the standard $\Lambda$CDM scenario is asymptotically recovered.

According to fig.~\ref{DeltaRhoPositive}, assuming a prior on $\Delta \rho_\Lambda$ and that the diffused energy is exclusively associated with black holes, the average energy (in cosmological time) transferred from black hole spin would correspond to an energy $E/V_{\rm Total}\approx1.69\times10^{9}\rm M_{\odot}/Mpc^3$ for the continuous model under the assumption of energy diffusion from matter into dark energy. Interestingly, according to recent estimates, the local ($z\approx0$) black hole mass density in early-type galaxies --- obtained from measurements of the velocity dispersion of early-type galaxies from the Sloan Digital Sky Survey --- is around 
$\rho_{\rm BH} (z=0)\sim (2.5 \pm 0.4) \times 10^5 \left(\frac{100\, h}{65}\right)^2 \, M_\odot/{\rm Mpc}^{3}.$ \cite{Yu:2002sq}. Thus, under this constraint, more than the total energy budget in the form of black holes would have to be dissipated through this diffusion mechanism. This is, in principle, problematic. However, such estimates are based on observations of black holes in nearby galaxies at low redshift. Significant contributions may arise, for example, from primordial black holes, whose formation history and population evolution are still not well understood, as well as from black holes that have already evaporated.

\section{Conclusions}
\label{sec:Conclusions}

In the cosmological context, an effective cosmological constant arises naturally from the non-conservation of the energy–momentum tensor in unimodular gravity. This apparent violation of energy conservation is expected to emerge naturally from the quantum nature of matter within the framework of objective collapse theories. Another possible manifestation arises from a black-hole granularity friction mechanism, in which black holes lose energy in the form of spin through interactions with the underlying granularity. This lost energy can be interpreted as an effective cosmological constant, exhibiting a natural dynamical behavior and providing a potential explanation for the origin of the cosmological constant.

If the evolution of $\rho_{\Lambda}$ is driven by a black-hole-related mechanism, for instance through friction-like effects, one would expect information about its behavior to be encoded in the cosmological density function, $f_{\rm BH}(M, J)$, which describes the abundance of black holes as a function of their mass and spin. In this framework, we extend the analysis of \cite{Landau:2022mhm} by parameterizing the diffusion of dark matter and baryonic matter into an effective dark energy component using a simple smooth step function. This approach is motivated by the absence of a well-defined cosmological density function, $f_{\rm BH}$, which would allow one to determine $\mathbf{J}(t)$, as well as by the lack of a clear consensus on how a diffusion mechanism arising from a quantum description of matter could manifest in the cosmological framework.

Still, in this context, contrary to the intuitive expectation of a diffusion process from the matter sector into dark energy, the results do not show a conclusive preference for either negative or positive values of $\Delta \rho_{\Lambda}$ in any of the scenarios explored. In principle, this is not problematic, as mechanisms capable of producing time decreasing cosmological constant may arise when considering, for example, quantum effects manifesting in a cosmological scenario \cite{Josset:2016vrq}.

A delicate aspect is related to the fact that, although the model performs better on average, once the data and the number of parameters are taken into account, no preference is found. In more general analyses, the transition can be characterized by a function more complex than the simple step described by eqs. \eqref{DE_1} and \eqref{DE_2}. According to eq. \eqref{eq. current}, contributions associated with different effects are expected to become relevant at different epochs, supporting a more general framework in which multiple transitions at different times may occur, or even motivating extensions beyond a simple monotonic behavior. For instance, the sign of $\mathbf{J}_{\rm CSL}$ may vary, as it is expected to decrease the effective cosmological constant. Exploring this scenario could provide valuable insights into the overall viability of this type of diffusion mechanism (see, for example, the expressions in eqs.~\eqref{rho_b general} and \eqref{rho_c general}). Additionally, in this analysis, we fix the summed neutrino mass to $\sum m_\nu = 0.06 \, \mathrm{eV}$. Alternatively, $\sum m_\nu$ can be treated as a free parameter in order to investigate possible degeneracies with the unimodular parameters.

\acknowledgments

This research was performed using the services and resources provided by the LAMOD-UNAM project through the Atocatl and Tochtli clusters. LAMOD is a collaborative effort between DGTIC and the research institutes of astronomy, nuclear science, and chemistry at UNAM. DS  acknowledges support from the Proyect PAPIIT  No IG100124,  and a sabbatical year Fellowship from UNAM,  through a  PASPA/DGAPA, and the hospitality of the Departament de Física Quantica i Astrofísica  Universitat de Barcelona. MV, SF acknowledge PAPIIT-IN116024 and PAPIIT-IN115424.

\bibliographystyle{JHEP}
\bibliography{bibliography}

\appendix
\section{Testing the agreement of observables individually for unimodular gravity}
\label{sec:Appendix A}

\begin{figure}[htbp]
  \centering
    \includegraphics[width=0.48\textwidth]{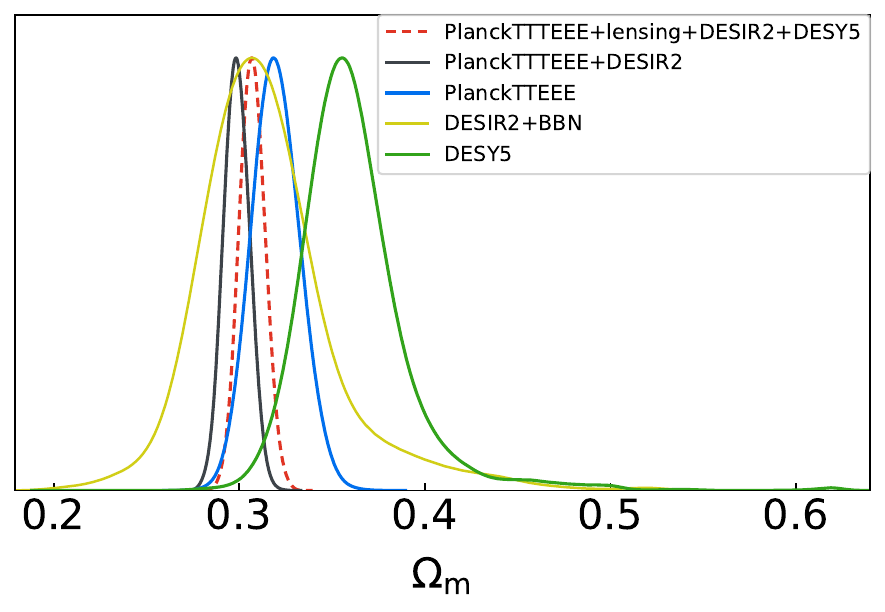}
    \includegraphics[width=0.48\textwidth]{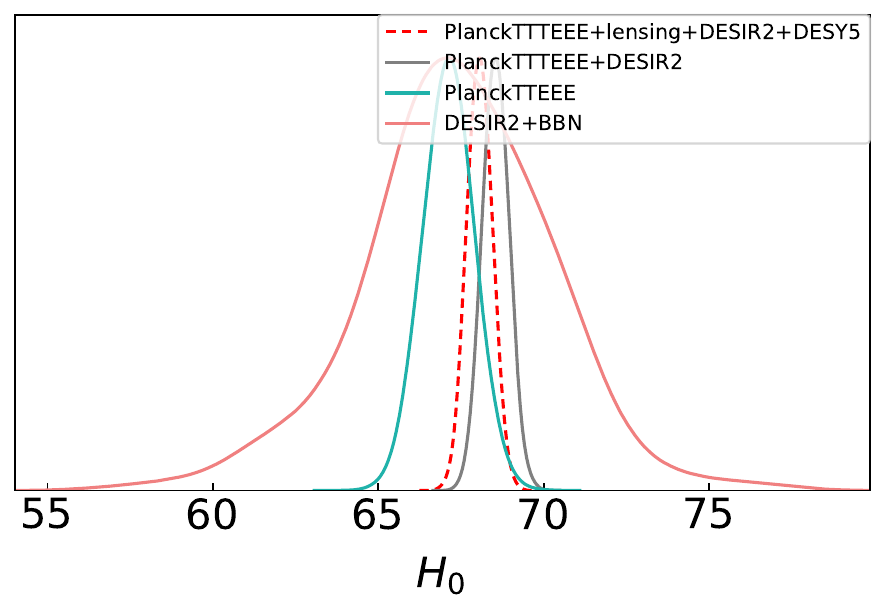}

        \caption{Posterior results for the $\Omega_m$ parameter using individual constraints for the continuous model, compared with the full combination corresponding to the fiducial model of the paper.}
    \label{Apendix_Omega_m}
\end{figure}

For this work, we combine CMB, supernova, and baryon acoustic oscillation data from Planck, DESY5, and Data Release 2 of the DESI collaboration. Hence, an individual analysis was required to validate the combination, ensuring that the constraints on the parameters subject to the largest tension are consistent at least at the $3\sigma$ level. Figure~\ref{Apendix_Omega_m} present the marginalized posterior of $\Omega_m$ and $H_0$ obtained from the Planck temperature and polarization power spectra, DESY5 supernova data, and DESI BAO data using Big Bang Nucleosynthesis prior on $\Omega_b h^2=0.02218\pm 0.00055$ derived from the PRyMordial code \cite{Schoneberg:2024ifp}. 

Modeling the distributions as one-dimensional Gaussian functions, we find that the tension between the supernova and BAO measurements is $0.92\sigma$ for $\Omega_m$ posterior distribution. That is, it lies within the standard $3\sigma$ level. Similarly, for $H_0$, we find a tension of $0.45\sigma$ between the CMB and BAO measurements.

\begin{figure}[htbp]
  \centering

    \includegraphics[width=0.49\textwidth]{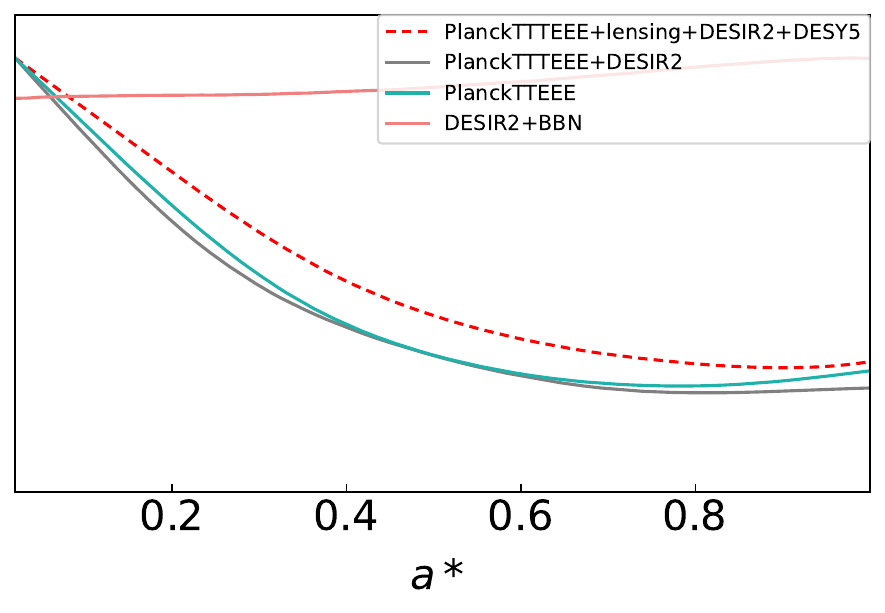}
    \includegraphics[width=0.49\textwidth]{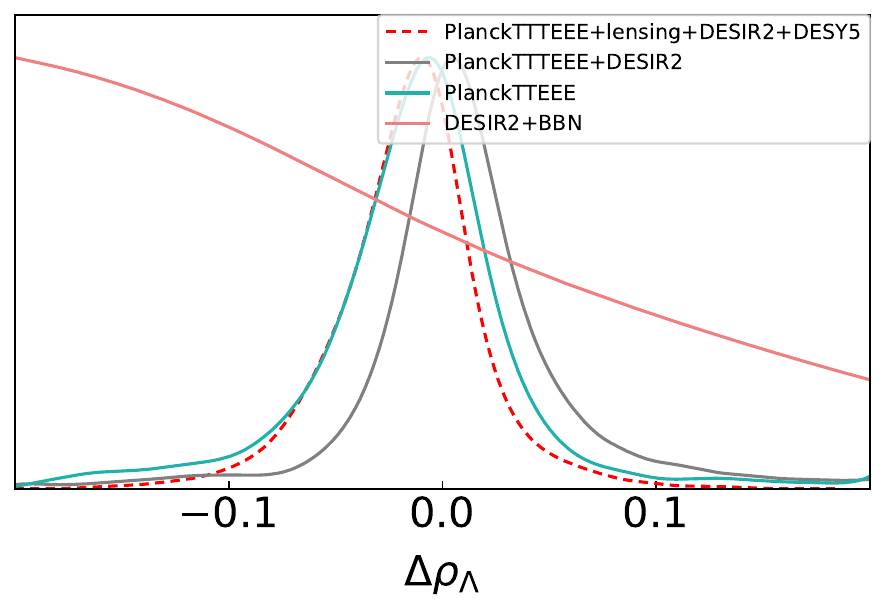}

    \includegraphics[width=0.49\textwidth]{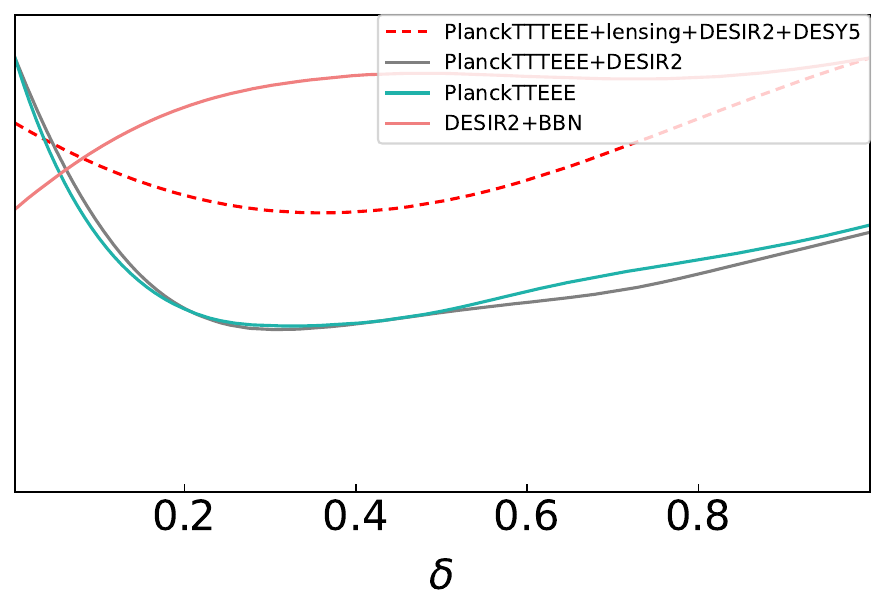}
        \caption{Posterior results for the continuous model parameters $\Delta \rho_{\Lambda}$, $a^*$ and $\delta$.}
    \label{Apendix_Delta_Rho_Lambda_a_start_delta}
\end{figure}

We can also examine the behavior of the unimodular parameters for each individual data set. Figures \ref{Apendix_Delta_Rho_Lambda_a_start_delta} illustrate the corresponding posterior distribution, with the dashed red line indicating the combined data set case. Planck CMB data show better agreement with a scenario in which there is diffusion from the dark energy into dark matter. Once the data are combined, the model tends to prefer, on average, negative values of $\Delta \rho_{\Lambda}$. Once the Planck data are combined with BAO information, a slight shift toward positive values of $\Delta \rho_{\Lambda}$ is observed. In particular, for the Planck dataset, $\Delta \rho_{\Lambda} = -0.01 \pm 0.05$, whereas the combination of Planck and DESI data yields $\Delta \rho_{\Lambda} = 0.01 \pm 0.05$. Estimates based on DESI DR2 with a prior on $\omega_b$, do not yield strong constraints on the model parameters.

\section{Dark energy parametrizations and the unimodular model}

\label{sec: Appendix B}

As we mentioned in previous sections, the parameterizations of the equation of state of dark energy models have been actively investigated. This is because the dark energy component is usually attributed to a matter entity whose motivation is primarily empirical. In this work, we do not adjust an equation of state since we are not thinking of this gravitational contribution as a new component; instead, we attribute it to a consequence of non-conservation of the energy-momentum tensor, which may be associated with an interaction between the granular spacetime structure and high-curvature entities (such as black holes) or the quantum nature of the matter.

For observables such as distances used in BAO and supernova analyses, the relevant quantities depend on $H(z)$ and its integral ($D_V/r_d$ and $D_M/r_d$ \cite{DESI:2025zgx}). This allows for a direct comparison with standard parametrizations through the contribution of $\rho_\Lambda$ in eq. \eqref{eq. Friedmann and cont}. By comparing the dark energy growth functions in the continuous and discrete models with those used in the most widely adopted parameterizations, such as CPL, BA and JBP.
The conventional $\omega_0$–$\omega_a$ parametrization fails to capture these models. This is the reason why a dedicated parameter inference analysis was required, and why a direct comparison with the standard commonly performed in this context of a state equation is not appropriate, particularly when late-time probes are used. For late-time observations, the situation is more subtle, since they distinguish between baryonic matter and dark matter, and eqs. \eqref{omegab}, \eqref{omegadm}, \eqref{rhb_C1}, and \eqref{rhdm_C2} clearly differ from the standard evolution, making a direct comparison impossible.

\end{document}